\documentclass[fleqn,usenatbib]{mnras}

\usepackage{newtxtext,newtxmath}
\usepackage[T1]{fontenc}
\usepackage{graphicx}
\usepackage{xcolor}
\usepackage{amsmath}
\usepackage{gensymb}
\usepackage{float}
\usepackage{natbib}
\usepackage{physics}
\usepackage{subcaption}

\DeclareRobustCommand{\VAN}[3]{#2}
\let\VANthebibliography\thebibliography
\def\thebibliography{\DeclareRobustCommand{\VAN}[3]{##3}\VANthebibliography}

\title[Sky Map Resolution Requirements for Signal Extraction]{Quantifying Sky Map Resolution Requirements for Beam Chromaticity Correction in Sky-Averaged 21 cm Experiments}

\author[Dragovic et al.]{
Aleksandra Dragovic,$^{1, 2}$\thanks{E-mail: ad2270@cam.ac.uk}
Dominic Anstey,$^{1, 2}$
Harry T. J. Bevins$^{1, 2}$
and Eloy de Lera Acedo$^{1, 2}$\\
$^{1}$Astrophysics Group, Cavendish Laboratory, University of Cambridge, J. J. Thomson Avenue, Cambridge, CB3 0US, UK\\
$^{2}$Kavli Institute for Cosmology in Cambridge, University of Cambridge, Madingley Road, Cambridge, CB3 0HA, UK\\
}

\date{Accepted XXX. Received YYY; in original form ZZZ}
\pubyear{\the\year{}}

\begin{document}
\label{firstpage}
\pagerange{\pageref{firstpage}--\pageref{lastpage}}
\maketitle

\begin{abstract}
The 21 cm hyperfine transition of neutral hydrogen provides one of the few direct probes of the early cosmic history. High-redshift detections of this signal could shine light on the poorly understood epochs of the Dark Ages and Cosmic Dawn. However, this signal is masked by bright foregrounds and is distorted by the chromaticity of the antenna used to detect it. Correcting for the chromaticity requires an accurate representation of the radio sky across relevant frequencies. However, base sky maps used for this correction, such as instances of the Global Sky Model (GSM), are limited by resolution, calibration and extrapolation uncertainties, especially when scaled to lower frequencies relevant for Cosmic Dawn studies. This work quantifies how accurately the base map must represent the true sky to produce a reliable beam correction. Using simulated sky data, we generate beam chromaticity corrections generated from progressively degraded versions of the base map, and compare how well they do in comparison to a correction based on the full-resolution map. We find that degradation in map resolution of $\sim 1^\circ$ does not introduce measurable increases in the amplitude of the residuals when correcting for the chromaticity and subtracting three different foreground models (two polynomials and a power law expansion). These results provide practical constraints on the required resolution and accuracy of sky models used in beam correction pipelines, informing future design and calibration strategies for global 21 cm experiments, and future efforts to map the low frequency sky.
\end{abstract}

\begin{keywords}
methods: data analysis – methods: statistical-cosmology: dark ages, reionization, first stars
\end{keywords}

\section{Introduction}
\label{sec:introduction}
 Reconstructing the timeline of the early Universe is one of the central goals of 21-cm cosmology. Optical or infrared tracers cannot directly probe the epochs of Dark Ages and Cosmic Dawn, leading to poor understanding of the era between recombination and the formation of the first luminous sources. The abundance of neutral hydrogen in the early Universe makes the 21-cm spectral line a powerful probe into this period, and tracing its' differential brightness temperature against the CMB can explain the formation of the first stars, black holes and galaxies. 
 In 2018, the EDGES (Experiment to Detect the Global Epoch of Reionization Signature) collaboration reported the detection of an absorption feature in the sky-averaged radio spectrum, centered at 78 MHz \citep{EDGES2017}, corresponding to a redshift of approximately $z=17$. The signal exhibited a flattened Gaussian profile with an amplitude of approximately $0.5^{+0.5}_{-0.2}$ K, significantly deeper than predicted by standard cosmological models (\citealt{Cohen_Fialkov_2017}; \citealt{Cohen_Fialkov_2020}). If cosmological, it implies that the intergalactic medium was colder than expected or that the radio background was substantially enhanced beyond the CMB \citep{FengHolder2018,Modeling_from_edges}. The absorption profile was recovered by fitting the foreground and the 21 cm signal simultaneously, using a weighted least squares solution. The adopted foreground model was a five term function motivated by the physics of Galactic synchrotron emission and ionospheric absorption/emission (a linear approximation to an underlying physically parameterized form). A separate, purely empirical polynomial model with a variable number of terms was also used in validation trials and gave consistent results.

\begin{figure*}
    \centering
    \includegraphics[width=\textwidth]{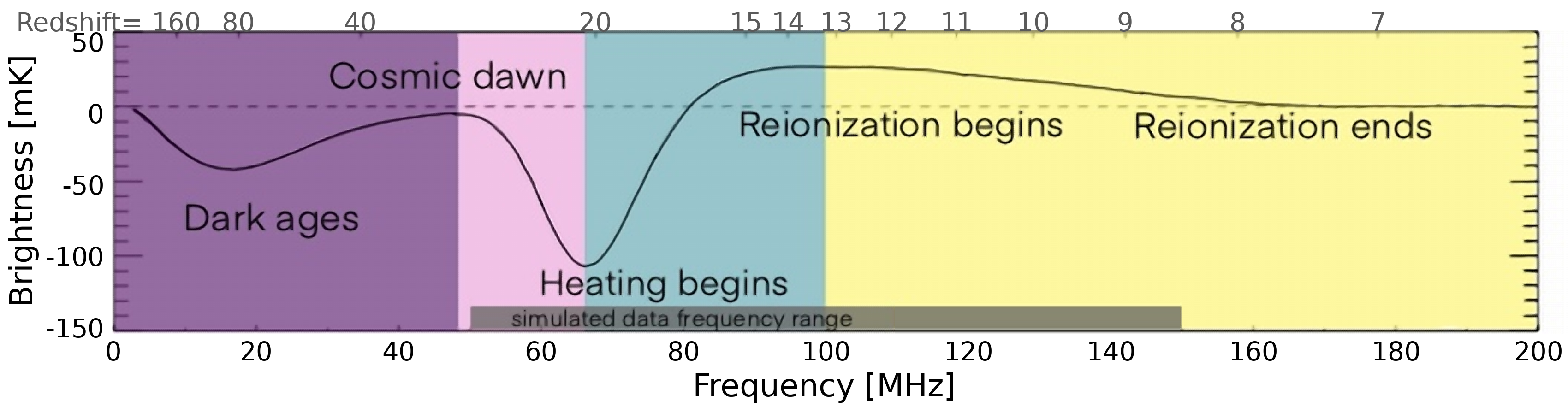}
    \caption{A model of the Global 21 cm line temperature change through main cosmic events (highlighted in different colours), also showing the frequency range simulated data in this paper uses. From left to right:
    era of collisional coupling (purple), onset of Lyman-$\alpha$ coupling (pink), onset of X ray heating (turqouise), and era of reionization (yellow).}
    \label{fig:cosmic_timeline}
\end{figure*}

The reported absorption profile, particularly its amplitude and shape, has been met with skepticism. A primary concern is that instrumental systematics or residual foreground contamination may mimic the cosmological signature, as shown by \citet{Pober_Sims_2019}, \citet{Hills_2018} and \citet{Singh_2019_sunisoidal_edges_feature}. \citealt{Pober_Sims_2019} and \citet{Maxsmoothpaper} also showed a presence of a damped sinusoidal systematic which is strongly preferred in the EDGES data. In response, several global 21 cm experiments made to detect the spatially averaged 21 cm signal have intensified efforts to verify or refute the claimed feature. The SARAS 3 experiment \citep{SARAS3} has reported upper limits incompatible with the amplitude of the EDGES signal and also suggested unaccounted systematics could be mistaken for the 21 cm trough in \citep{SARAS_disputing_EDGES}. One ongoing project aiming to verify this is REACH (Radio Experiment for the Analysis of Cosmic Hydrogen; \citealt{REACH}), which will employ several antennas with significantly different spectral responses for robust systematics suppression, a Bayesian inference pipeline for signal extraction, and forward modelling approach based on physical models of the foreground (\citet{REACH_bayesian_pipeline_dominic}, \citet{Bayesian_calibration_2023}, \citet{Jacobs_pipeline_paper_2026}). REACH aims to determine whether the currently accepted astrophysical model of the Universe has to be revised by, for instance, adding an excess radio background or extra cooling at high redshifts. 

One of the specific systematics that any such pipeline must contend with is beam chromaticity: because a real antenna's sensitivity pattern varies with frequency, it imprints non smooth spectral structure onto an otherwise smooth foreground, which can mimic or mask the very signal these experiments are trying to recover. The standard way to remove this is a beam chromaticity correction, computed from an external sky map and the known antenna beam (\S\ref{sec:skymaps_correction}); this correction was central to the original EDGES analysis \citep{EDGES2017, Mozdzen2016}, and is the technique we adopt and test here. Existing low-frequency sky maps, however, are of limited resolution and are often themselves derived from higher frequency surveys such as the Haslam 408\,MHz map which has an angular resolution of $56\arcmin$ \citep{Haslam}. This raises a practical question that, to our knowledge, has not been directly quantified: how accurate does the sky map used for beam chromaticity correction need to be for the correction to remain reliable?
 
In this paper we answer this question using simulated data for testing, which will be described in detail in Section \ref{data_simulation}. We systematically degrade the resolution of the base sky map used to generate the chromaticity correction that we later apply to the simulated data sets, and quantify the resulting change in fit residuals and recovered signal quality relative to a full resolution correction, across three foreground parameterisations and two fitting frameworks (constrained and unconstrained). This method is 
applicable to any sky averaged 21 cm experiment relying on a beam chromaticity correction of this kind.

\begin{figure*}
    \centering
    \includegraphics[width=\textwidth]{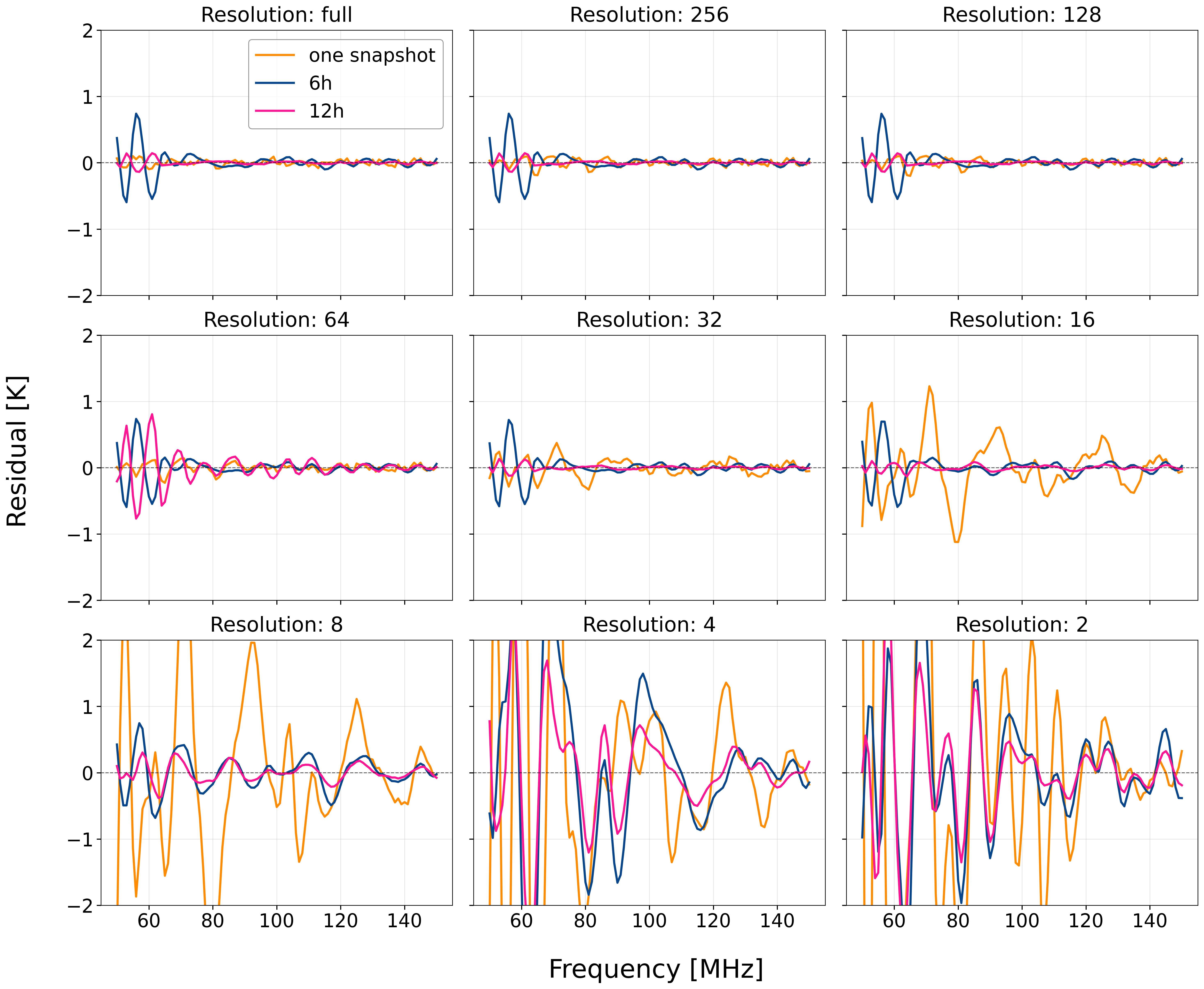}
    \caption{Chromatic residuals of three different data sets, all simulated using Eq.~\ref{eqn:Tdata}. The difference between the data sets are time durations described in \S\ref{data_simulation} (12h and 6h time integrated data sets, and one single snapshot data set). Data was fit with a constrained polynomial, after applying a chromaticity correction given in Eq.~\ref{eqn:beamcorrection} where resolution-degraded versions of the sky map were used.}
    \label{fig:all_gal_times_nosignal}
\end{figure*}

\subsection{Background}
\label{sec:background}
During the Cosmic Dawn, Lyman-$\alpha$ photons from the first stars caused coupling between the spin temperature $T_{\mathrm{s}}$ and the kinetic temperature of the hydrogen gas $T_{\mathrm{K}}$ through the Wouthuysen-Field effect \citep{Wouthuysen, Field_a, Field_b}. This coupling produces a global absorption feature in the radio spectrum as seen in Figure~\ref{fig:cosmic_timeline}. As X-rays from the first exotic objects began to heat the intergalactic medium, $T_{\mathrm{K}}$ (and hence $T_{\mathrm{s}}$) rises, reducing the depth of the absorption trough and potentially leading to 21\,cm emission. Finally, as reionization progresses and the neutral fraction $x_{\mathrm{H}}$ declines toward zero, the 21\,cm signal vanishes entirely. This sequence of absorption, heating, and ionization encodes rich information about the formation of the first luminous sources and constrains the Cosmic Dawn.
Detecting the redshifted 21 cm line from Cosmic Dawn remains one of the most technically challenging tasks in observational cosmology. The foregrounds are several orders of magnitude stronger than the signal’s expected brightness temperature. Frequency dependent antenna beam effects, calibration uncertainties and RFI are all present as well.
Foregrounds at these low radio frequencies are dominated by Galactic synchrotron emission and unresolved extragalactic sources, which are several orders of magnitude brighter than the anticipated 21 cm signal. Foreground modeling and removal continues to be a focal point of research \citep[e.g.]{Dominic2023,Pagano_bayesian_modeling}. The use of high order foreground models for spectral fitting, including the five term physically motivated model employed by EDGES, has been noted to potentially either remove components of the cosmological signal or introduce artifacts \citep{Singh_2019_sunisoidal_edges_feature, Hills_2018}. Alternatives, such as maximally smooth functions \citep{Maxsmoothpaper, msf_proposition_2015, msf_proposition_2017}, or physically motivated foreground models \citep{Dominic2021}, have been proposed to better constrain the spectral structure of foregrounds without overfitting or introducing artificial features.

The remainder of the paper is organised as follows: \S\ref{sec:methods} describes the simulated data used, the foreground models tested, the beam chromaticity correction and resolution-degradation procedure;
\S\ref{sec:constrained} and \S\ref{sec:unconstrained} present the constrained (\textsc{maxsmooth}) and unconstrained (Bayesian nested sampling) fitting methods; \S\ref{sec:results} presents our main results on residuals and signal recovery as a function of map resolution; \S\ref{sec:conclusion} concludes.

\section{Methods}
\label{sec:methods}
 
\subsection{Data Simulation}
\label{data_simulation}
We simulate data using the following equation:
\begin{equation}
    \label{eqn:Tdata}
    T_{\text{model}}(\nu) = \int D(\Omega, \nu)\, T_{\rm sky}(\Omega, \nu)\, d\Omega + \mu(\nu),
\end{equation}
where the known dipole antenna beam model $D(\Omega, \nu)$ is convolved with a sky map $T_{\rm sky}(\Omega, \nu)$ (\S\ref{sec:skymaps_correction}) and Gaussian
noise $\mu(\nu)$ is added. We use this simple foreground model, without ionospheric or polarization terms \citep{emma_ionospheric_effects}, to isolate the chromatic systematics that are the focus of this work.

As the Galactic plane moves relative to the beam over the course of an observation, it modulates the chromatic ripples that partially average down with longer integration. To quantify this, we generate three data sets, all starting 2022-05-01 00:00:00 (UTC): a single snapshot with the Galactic disc above the horizon, and 6 hour and 12 hour time integrated observations. When looking at the top left plot in Figure~\ref{fig:all_gal_times_nosignal} for example, integrating over a prolonged period of time (12 hours) as the sky rotates, shows residuals comparable to residuals of a one snapshot. The spectral structures average out, so a correction that assumes a uniform spectral index will result in low amplitude residuals. The other data set integrated over a time period of 6 hours shows $\sim 0.5\,\mathrm{K}$ increase in residual amplitude compared to the other two data sets. This indicates that partial integration might be worse than none. A single snapshot shows a fixed sky image, and whatever spectral structure it has, correcting for beam chromaticity and fitting a constrained polynomial can absorb it cleanly. It has also been shown that the performance of the chromaticity correction we use in this paper, described in \S\ref{sec:chromaticity}, is LST dependent \citep{Mozdzen2016, Mozdzen2017}.
These residuals were obtained by fitting data using methods later described in Section \S\ref{sec:constrained} using a log-log polynomial in Eq. (\ref{eqn:loglog_polynomial_intro}) to model and subtract the foreground.

\subsection{Foreground models}
\label{sec:foreground_modelling}
Because Galactic and extragalactic synchrotron foregrounds couple with the frequency dependent antenna beam, their intrinsically smooth power-law spectrum
acquires non-smooth chromatic distortions comparable in amplitude to the 21 cm signal itself (see \S\ref{sec:skymaps_correction} for the beam correction we
apply). We test three foreground parameterisations spanning empirical polynomial forms and the physically motivated power law form used by EDGES \citep{EDGES2017, Modeling_from_edges} to check whether our conclusions about sky map resolution requirements are robust to this choice, rather than an artifact of one particular parameterisation.
It is important to note that the EDGES power law expansion model has been noted to potentially absorb components of the cosmological signal or introducing artifacts \citep{Hills_2018, Singh_2019_sunisoidal_edges_feature, Pober_Sims_2019}.
Alternatives such as Maximally Smooth Functions \citep{Maxsmoothpaper} or physically motivated foreground models have been proposed to better constrain the spectral structure of the foreground without overfitting. 

The first model we use is a difference polynomial,
\begin{equation}
    \label{eqn:differencepolynomial_intro}
    y = \sum_{k=0}^{N} a_k \left( \nu - \nu_0 \right)^{k},
\end{equation}
where $\nu_0 = 100\,\mathrm{MHz}$ is a pivot frequency taken to be the midpoint of the observed band and $a_k$ are free model parameters. The pivot point reduces correlations between polynomial coefficients and improves numerical conditioning. This is a general-purpose parameterisation that makes no strong assumptions about the spectral shape of the foreground.

The second is a log-log polynomial,
\begin{equation}
    \label{eqn:loglog_polynomial_intro}
    \log_{10}(y)={\displaystyle\sum_{k=0}^{N} a_k 
    \left[\log_{10}\!\left(\frac{\nu}{\nu_0}\right)\right]^k}.
\end{equation}
 This parameterisation is physically motivated by the power-law behaviour of synchrotron foregrounds, which appear as straight lines in log-log space. Working in logarithmic coordinates aligns the basis functions with the natural spectral structure of the foreground, meaning fewer terms may be needed to achieve a given quality of fit.

The third is the EDGES style power-law expansion,
\begin{equation}
    \label{eqn:edges_powerlaw_intro}
    T(\nu) = \sum_{k=0}^{N} a_k \, \nu^{\,k - 2.5},
\end{equation}
where the exponent offset of $-2.5$ \citep{Mozdzen2016} reflects the characteristic synchrotron spectral index of diffuse Galactic emission \citep{EDGES2017}. This parameterisation was used in the original EDGES signal detection and is included here to allow direct comparison with that analysis.
Each parameterisation is fit under two frameworks: a constrained (\S\ref{sec:constrained}) and an unconstrained approach (\S\ref{sec:unconstrained}). Comparing the two lets us distinguish residuals that are robust across fitting frameworks and functional forms and therefore more likely to reflect genuine chromatic structure or a recovered signal from those sensitive to the modelling choice, which would indicate a systematic bias in the analysis.

\subsection{Beam chromaticity correction and resolution degradation}
\label{sec:skymaps_correction}

Global sky maps at frequencies at which REACH operates are not directly available, and would in any case likely already contain contributions from the global 21\,cm signal itself. Higher frequency sky maps are instead used and scaled to the target frequency range. In this work, the base sky map $T_{\mathrm{base}}(\theta,\phi,t)$ used to generate the chromaticity correction is an instance of the Global Sky Model (GSM) at 230\,MHz \citep{deOliveira-Costa_GSM2008}, one of the most widely used sky models in radio astronomy. The GSM map is an amalgam of many different maps, including the Haslam all-sky map \citep{Haslam}, which is publicly available and also one of frequently used all-sky maps in radio astronomy.\footnote{\url{https://lambda.gsfc.nasa.gov/product/foreground/fg_2014_haslam_408_info.html}} To enable frequency scaling, we constructed a spectral index map by calculating, for each pixel, the index $\beta(\theta,\phi)$ needed to map GSM's pixel values at 408\,MHz to the corresponding pixel values at 230\,MHz \citep{Dominic2021}:
\begin{equation}
    \label{eqn:spectralindex}
    \beta(\theta, \phi) = 
    -\frac{
        \log\left(\frac{T_{230}(\theta, \phi)-T_{\mathrm{CMB}}}
        {T_{408}(\theta, \phi)-T_{\mathrm{CMB}}}\right)
    }{
        \log\left(\frac{230}{408}\right)
    },
\end{equation}
where $T_{408}$ is the Haslam (408 MHz) all-sky map brightness temperature per pixel, $T_{230}$ is the GSM (230 MHz) brightness temperature per pixel.

For each of the time-integration cases described in Section~\ref{data_simulation}, two different approaches are used to scale the sky map from the reference frequency to the observation frequencies. The first assumes a spatially uniform spectral index across the sky, such that all pixels follow the same power law frequency dependence. The sky temperature at 
each frequency is then given by
\begin{equation}
    \label{eqn:skytemp}
    T_{\mathrm{s}}(\theta,\phi,\nu,t) = 
    \left[ T_{\mathrm{base}}(\theta,\phi,t) - T_{\mathrm{CMB}} \right]
    \left( \frac{\nu}{\nu_{\mathrm{base}}} \right)^{-2.5} + T_{\mathrm{CMB}},
\end{equation}
uniformly scaled to a reference frequency of $\nu_{\mathrm{base}} = 79\,\mathrm{MHz}$ with a fixed spectral index of $\beta = -2.5$.
$T_{\mathrm{CMB}}$ is the CMB temperature of 2.725\,K. Such a correction was previously applied to the EDGES data in \citet{Mozdzen2016}. The second approach allows the spectral index to vary across the sky, scaling each pixel independently using the full spectral index map. The sky temperature 
is then
\begin{equation}
    \label{eqn:skytemp_full}
    T_{\mathrm{sky}}(\theta,\phi,\nu,t) = 
    \left[ T_{\mathrm{base}}(\theta,\phi,t) - T_{\mathrm{CMB}} \right]
    \left( \frac{\nu}{\nu_{\mathrm{base}}} \right)^{\beta(\theta,\phi)} + T_{\mathrm{CMB}} ,
\end{equation}
where $\beta(\theta,\phi)$ is the spatially varying spectral index from Eq.~\ref{eqn:spectralindex}, applied independently to each pixel.

\begin{figure*}
    \centering
    \includegraphics[width=\textwidth]{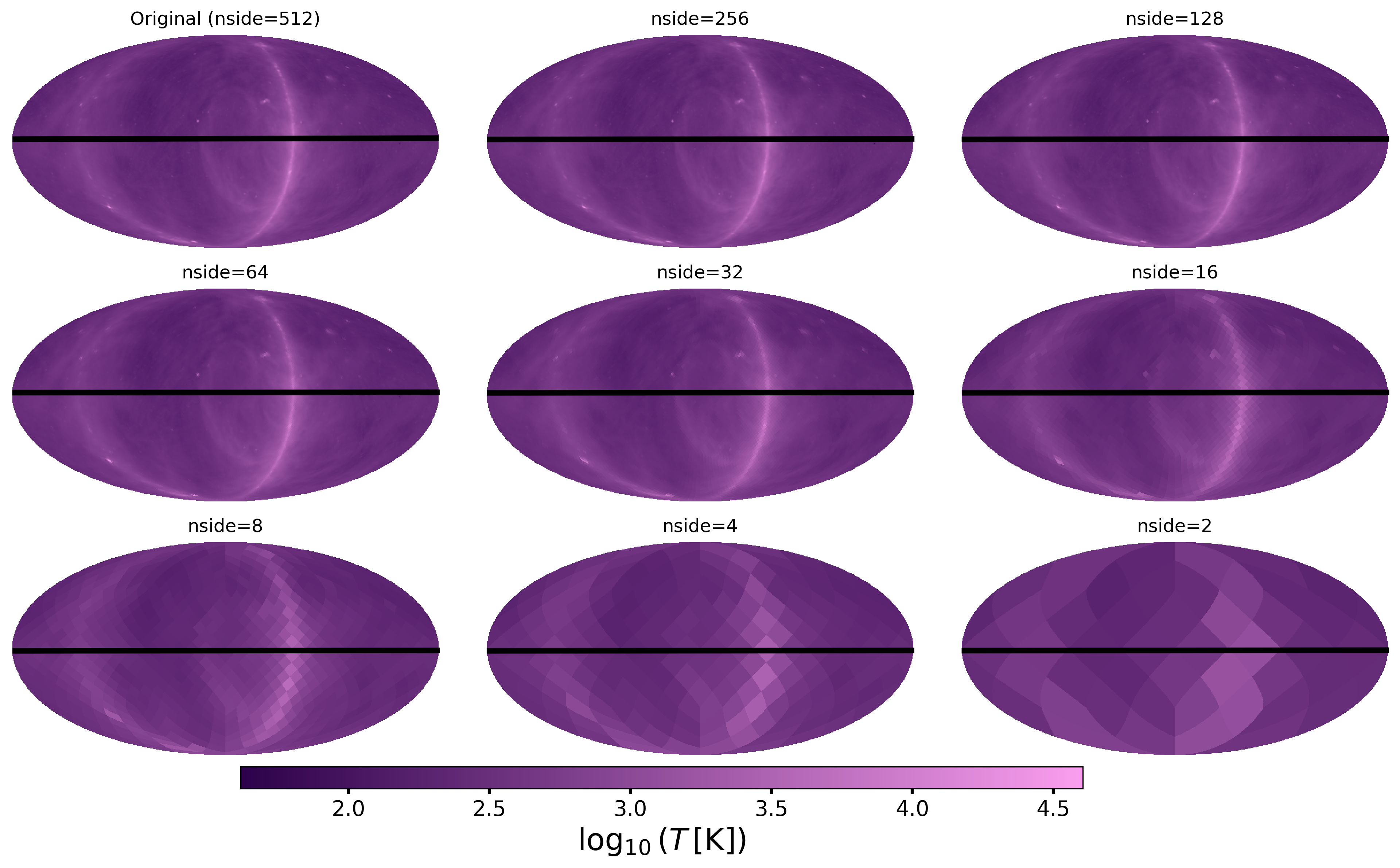}
    \caption{Example of downgraded sky maps across multiple resolutions. The horizon is shown as a black line; regions above it are visible to the telescope. The first plot (upper left) shows the brightness temperature variations across the sky. The bright streak is the Galactic disc. The error of each blurred map is shown in Table \ref{tab:rms}, as RMS of the difference between the original map and each blurred one.}
    \label{fig:downgraded_maps}
\end{figure*}

\subsection{Chromaticity correction}
\label{sec:chromaticity}
In an ideal scenario, an antenna beam would be entirely independent of frequency, with its sensitivity pattern remaining constant across all observed frequencies. However, any wide band antenna exhibits frequency dependent chromatic effects that introduce spectral structure into the observed signal. For example, at higher frequencies, above 100\,MHz, the dipole beam used by REACH \citep{REACH_antenna_design} begins to split, and its maximum sensitivity no longer coincides with the zenith. These chromatic distortions, combined with the intrinsic spectral variations of the sky, can obscure or entirely mask the faint 21\,cm signal. To correct for them, we can apply a beam chromaticity correction factor 
$B_{\mathrm{factor}}(\nu, t)$, defined as
\begin{equation}
    \label{eqn:beamcorrection}
    B_{\mathrm{factor}}(\nu,t) =
    \frac{
        \int_0^{4\pi}
        \left[ T_{\mathrm{s}}(\theta,\phi,\nu_{\mathrm{r}},t)  \right] 
        D(\theta,\phi,\nu) \, \mathrm{d}\Omega
    }{
        \int_0^{4\pi}
        \left[ T_{\mathrm{s}}(\theta,\phi,\nu_{\mathrm{r}},t)  \right] 
        D(\theta,\phi,\nu_{\mathrm{r}}) \, \mathrm{d}\Omega
    },
\end{equation}
where the numerator represents the beam-weighted sky temperature at frequency $\nu$, integrated over all solid angles, and the denominator is the same quantity evaluated at the reference frequency $\nu_{\mathrm{r}}$ only. Note that the sky is held at a fixed frequency in this correction. This chromaticity correction was used on EDGES data in, for example \citet{Mozdzen2016} and later on simulated data sets by the REACH pipeline in \citet{Dominic2021}.
For this analysis, we took the beam $D(\theta,\phi,\nu)$ to be the REACH dipole beam. The corrected sky temperature is obtained by dividing the measured or simulated signal by $B_{\mathrm{factor}}$, removing the frequency dependent weighting introduced by the chromatic beam response.

A subtlety arises when this correction is applied in the context of a joint foreground and signal extraction pipeline, concerning the order of operations when injecting the cosmological signal relative to the beam chromaticity correction. When the signal is injected into the sky temperature before the correction factor $B_\mathrm{factor}(\nu)$ is applied, as is physically motivated since the signal originates in the sky before any instrumental response acts upon it, the corrected data takes the form
\begin{equation}
    T_\mathrm{corr}(\nu) = \frac{T_\mathrm{fg}(\nu) + T_{21}(\nu) - T_\mathrm{CMB}}{B_\mathrm{factor}(\nu)} + T_\mathrm{CMB},
\end{equation}
meaning that the signal present in the corrected data is $T_{21}(\nu) /  B_\mathrm{factor}(\nu)$ rather than $T_{21}(\nu)$ itself. Strictly speaking, this 
requires that the signal model in the likelihood be divided by $B_\mathrm{factor}(\nu)$ to match what is actually present in the data. However, in practice this correction introduces no measurable bias in the recovered signal parameters. This is because 
$B_\mathrm{factor}(\nu_\mathrm{r}) = 1$ by construction at the reference frequency $\nu_\mathrm{r} = 79\,\mathrm{MHz}$, which lies close to the centre of both injected signals considered in this work. Since $B_\mathrm{factor}$ is a small factor and remains close to unity near the reference frequency, the distortion it introduces to the signal shape is negligible, but it stays a useful correction for the foreground that is several orders of magnitude stronger than the signal. We verified this explicitly using controlled simulations on synthetic data, where joint foreground and signal recovery was shown to be unbiased regardless of whether the $B_\mathrm{factor}$ correction was applied to the signal term in the likelihood, or if the signal is injected into the data after correcting for chromaticity. Nevertheless, even though it is a small correction, we take it into consideration in the likelihood used for our nested sampling fits (Section~\ref{sec:blackjax}), ensuring that the signal template is consistent with the signal as it appears in the chromaticity-corrected data.

\begin{figure*}
    \centering
    \begin{subfigure}[t]{0.48\textwidth}
        \centering
        \includegraphics[width=\linewidth]{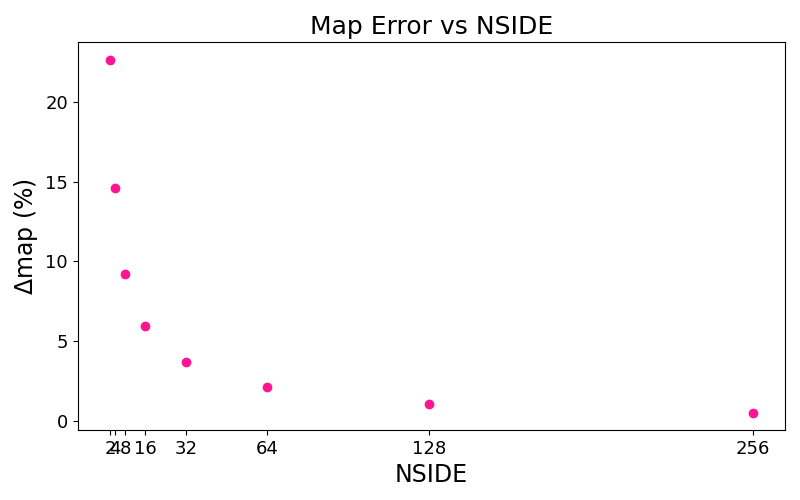}
        \caption{}
        \label{fig:error_map_percent}
    \end{subfigure}
    \hfill
    \begin{subfigure}[t]{0.48\textwidth}
        \centering
        \includegraphics[width=\linewidth]{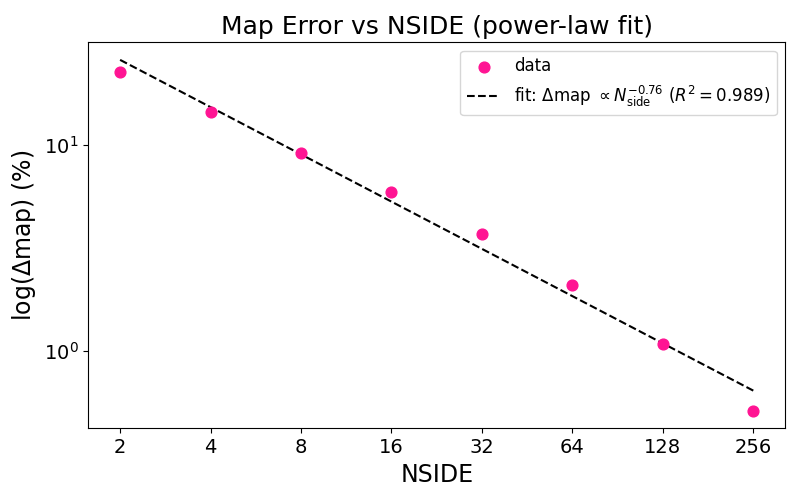}
        \caption{}
        \label{fig:error_map_percent_fit}
    \end{subfigure}
    \caption{(a) Mean relative error of each degraded map, calculated using Eq.~\ref{eqn:diffpercent} against its $N_{\rm side}$. (b) Map error as a function of $N_{\rm side}$ in log-space, against a power-law fit $\Delta_{\%}\mathrm{map} \propto N_{\rm side}^{-\alpha}$, confirming that the degradation error follows a power law.}
    \label{fig:error_map_combined}
\end{figure*}

The central question of this work is how well the base sky map used to compute $B_{\mathrm{factor}}$ needs to represent the true sky in order to produce a reliable correction. We quantify this by generating multiple chromaticity corrections, each computed 
from a progressively more degraded version of the base sky map, which are visualised in Figure~\ref{fig:downgraded_maps}, and testing these corrections on simulated data generated from the full-resolution map. We made use of the fact that the sky maps are stored in a \textsc{HEALPix} format \citep{healpix} \footnote{\url{https://healpix.sourceforge.io}}, the number of pixels in a \textsc{HEALPix} map is related to the $N_{\rm side}$ parameter by
\begin{equation}
\label{eqn:NSIDE}
N_{\text{pix}} = 12\,\mathrm{N_{\rm side}}^2,
\end{equation}
and the solid angle subtended by each pixel is
\begin{equation}
\Omega_{\text{pix}} = \frac{4\pi}{12\,\mathrm{N_{\rm side}^2}}.
\end{equation}
\textsc{HEALPix} has a hierarchical pixel structure in which pixel indices subdivide like a tree: one pixel at $N_{\rm side}=1$ becomes four pixels at $N_{\rm side}=2$, sixteen at $N_{\rm side}=4$, and so 
on, such that the mean value of the map is preserved under degradation. We degrade the GSM~230\,MHz map to $N_{\rm side} \in \{256, 128, 64, 32, 16, 8, 4, 2\}$, as illustrated in Figure~\ref{fig:downgraded_maps}.
Map degradation does not necessarily introduce non-smooth spectral structure, a spatially non-smooth sky map can still be spectrally smooth. However, noting that the beam $D(\theta,\phi,\nu)$ itself is frequency dependent and resolution degradation changes the sky brightness unevenly across the sky, this results in an error being introduced into $B_{\mathrm{factor}}(\nu)$.
Error introduced by the resolution degradation is calculated as pixel-by-pixel difference between the original base map and each degraded/blurred map, which we will refer to as map error. We quantify this map error in Table \ref{tab:rms}: first, the mean absolute deviation in Kelvin,
\begin{equation}
    \label{eqn:diffK}
    \Delta_{K}{\rm map}(N_{\rm side}) = \frac{1}{N_{\text{pix}}} \sum_{i=1}^{N_{\text{pix}}} 
    \left| T_{\mathrm{nside}}(\theta_i,\phi_i; N_{\rm side}) - T_{\mathrm{base}}(\theta_i,\phi_i) \right|,
\end{equation}
where $T_{\mathrm{base}}(\theta,\phi)$ is the original map at $N_{\rm side}=512$ and $T_{\mathrm{nside}}(\theta,\phi; N_{\rm side})$ is each of the blurred maps in $N_{\rm side} \in \{256, 128, 64, 32, 16, 8, 4, 2\}$. Second, the same quantity is expressed as a mean relative error in percentage,
\begin{equation}
    \label{eqn:diffpercent}
    \Delta_{\%}{\rm map}(N_{\rm side}) = \frac{1}{N_{\text{pix}}} \sum_{i=1}^{N_{\text{pix}}} 
    \left| \frac{T_{\mathrm{nside}}(\theta_i,\phi_i; N_{\rm side}) - T_{\mathrm{base}}(\theta_i,\phi_i)}{T_{\mathrm{base}}(\theta_i,\phi_i)} \right| \times 100.
\end{equation}
We visualise $\Delta_{\%}\mathrm {map}$ against $N_{\rm side}$ in Figure \ref{fig:error_map_percent} to show the trend it follows, which we show is almost a perfect power-law in Figure \ref{fig:error_map_percent_fit}.
In log-space we introduce $R^2$ to quantify how well the fit explains the variance in data:
\begin{equation}
    \label{eqn:rsquared}
    R^2 = 1 - \frac{\displaystyle\sum_i \Big[ \log \Delta_{\%}\mathrm{map}(N_{{\rm side},i}) - \big( \log C - \alpha \log N_{{\rm side},i} \big) \Big]^2}{\displaystyle\sum_i \Big[ \log \Delta_{\%}\mathrm{map}(N_{{\rm side},i}) - \overline{\log \Delta_{\%}\mathrm{map}} \Big]^2}.
\end{equation}
For a true power-law $R^2=1$. We find $R^2=0.989$, showing that the error increases almost as a perfect power-law as the map is degraded.
At lower $N_{\rm side}$ values, small-scale structures are lost as the map is averaged, bright features are smeared and sharp gradients are suppressed. This, combined with the frequency dependent beam pattern, introduces incorrect weighting into the beam convolution of 
Eq.~\ref{eqn:beamcorrection}.

\begin{table}
\centering
\begin{tabular}{c c c c}
\hline
NSIDE & $\Delta_K$ map (K) & $\Delta_{\%}$ map (\%) & $\Omega_{\rm pix}$ (deg) \\
\hline
256 & 1.070157e+00 & 0.5152 & 0.229 \\
128 & 2.268190e+00 & 1.0840 & 0.458 \\
64 & 4.474640e+00 & 2.1095 & 0.916 \\
32 & 8.025699e+00 & 3.7119 & 1.832 \\
16 & 1.299366e+01 & 5.9196 & 3.664 \\
8 & 1.962123e+01 & 9.1852 & 7.329 \\
4 & 2.946180e+01 & 14.5756 & 14.658 \\
2 & 4.075369e+01 & 22.6382 & 29.316 \\
\hline
\end{tabular}
\caption{Pixel-by-pixel difference of each degraded map $(T_{\mathrm{nside}} - T_{\mathrm{base}})$ compared to the original, expressed in units of Kelvin (Eq.~\ref{eqn:diffK}) and as a percentage (Eq.~\ref{eqn:diffpercent}).}
\label{tab:rms}
\end{table}

\subsection{Signal selection}
\label{sec:signal_selection}

We test signal recovery against two injected profiles visualised in Figure~\ref{fig:injectedsignals}: a standard astrophysical Gaussian, consistent with the Wouthuysen-Field coupling and X-ray heating history described in \S\ref{sec:background}, and the flattened Gaussian reported by EDGES \citep{EDGES2017} (Table~\ref{tab:example}), whose amplitude exceeds
standard $\Lambda$CDM predictions by a factor of $\sim$2.5 and whose flattened shape has no established astrophysical derivation \citep[see \S\ref{sec:introduction} for the proposed explanations, e.g.][]{excess_radio_background_fialkov, primordial_black_holes, signatures_of_cr_heating_tomas}.
Both signals have an amplitude comparable to the non-smooth chromatic distortions introduced by the beam (\S\ref{sec:skymaps_correction}), which is what makes their recovery challenging.

\begin{table}
\centering
\begin{tabular}{cccc}
\hline
Amplitude (A) & FWHM (w) & Center Frequency ($\nu_0$) & Flattening Factor ($\tau$) \\
\hline
0.155 K & 15 MHz & 85 MHz & 0 \\
\hline
0.5 K & 18.7 MHz & 78 MHz & 7 \\
\hline
\end{tabular}
\caption{Parameters of the signals used in data analysis.}
\label{tab:example}
\end{table}

\section{Constrained function fits}
\label{sec:constrained}

We fit the constrained models using \textsc{maxsmooth} \citep{Maxsmoothpaper}, which enforces a Maximally Smooth Function (MSF) form, requiring each higher order derivative to not change sign over the frequency range \citep{MSF_2015, MSF_2017},
In practice, this requires that the $m^{\mathrm{th}}$ derivative satisfies
\begin{equation}
    \label{eqn:derivative_constrains}
    \dv[m]{y}{x} \leq 0 \ \mathrm{or} \ \dv[m]{y}{x} \geq 0,
\end{equation}
for all $x$, enforcing a globally smooth functional form. \textsc{maxsmooth}'s sign-navigation algorithm efficiently searches the $2^{N-1}$ possible sign combinations to find the overall best fit \citep[see ][ for details]{Maxsmoothpaper}.

We use two of \textsc{maxsmooth}'s built-in bases the difference polynomial and log-log polynomial (Eqs.~\ref{eqn:differencepolynomial_intro}, \ref{eqn:loglog_polynomial_intro}) plus the EDGES style power-law expansion (Eq.~\ref{eqn:edges_powerlaw_intro}), implemented as a custom basis with the same MSF constraint applied to its $m^{\mathrm{th}}$ derivative:
\begin{equation}
    \label{eqn:edges_derivative}
    \dv[m]{T}{\nu} = \sum_{k=0}^{N} a_k \,
    \left[ \prod_{j=0}^{m-1} (k - 2.5 - j) \right]
    \nu^{\,k - 2.5 - m}.
\end{equation}

For a given parameterisation, \textsc{maxsmooth} determines the best-fit model by minimizing the $\chi^2$ statistic,
\begin{equation}
\label{chi_squared}
\chi^2 = \sum_{k}^{D} \left(y_k - y_{\mathrm{fit},k}\right)^2,
\end{equation}
subject to the smoothness constraints. The algorithm searches the parameter space for solutions that satisfy Eq.~\ref{eqn:derivative_constrains} and iteratively converges toward the global minimum. The use of multiple parameterisations allows us to assess whether any recovered signal residuals are robust to the choice of foreground basis, and to identify cases where the results are sensitive to the assumed functional form.

\section{Non-constrained function fits}
\label{sec:unconstrained}

To assess the impact of smoothness constraints on foreground modelling, we also fit the chromaticity corrected spectra with a unconstrained versions of the models that impose no restriction on the fitted function’s derivatives, in contrast to \textsc{maxsmooth} (\S\ref{sec:constrained}). We used a Bayesian nested sampling algorithm to perform these fits.

\subsection{Bayesian nested sampling}
\label{sec:nested_sampling}
 
The unconstrained fits use nested sampling \citep{nested_sampling}, which estimates posterior samples and the Bayesian evidence $\mathcal{Z}$:
\begin{equation}
\mathcal{Z} = \int \mathcal{L}(\boldsymbol{\theta})\,\pi(\boldsymbol{\theta})\,\mathrm{d}\boldsymbol{\theta},
\end{equation}
where $\mathcal{L}(\boldsymbol{\theta})$ is the likelihood and $\pi(\boldsymbol{\theta})$ is the prior distribution over model parameters $\boldsymbol{\theta}$.
It is well suited to the multimodal, correlated posteriors produced by polynomial foreground models. We use this evidence to perform model comparison between
foreground parameterisations and polynomial orders via Bayes factors.

\subsection{Implementation with BlackJAX}
\label{sec:blackjax}
We perform nested sampling inference using BlackJAX, \citep{Cabezas2024, nested_sampling_yallup_gpu}, a library for Bayesian inference implemented in JAX \citep{jax2018github}. JAX allows for more efficient evaluation of likelihoods and gradients. BlackJAX provides a range of sampling algorithms, including nested sampling, which we use here to explore the parameter space and obtain posterior distributions. The sampler operates on a log density function and requires only the likelihood and prior to be specified.
In this work, nested sampling is used to fit the foreground models without imposing constraints, enabling direct comparison with the constrained \textsc{maxsmooth} approach. The method yields weighted posterior samples, which are used to construct model averages and quantify uncertainties. At high number of polynomial terms N, parameter space is highly multi-modal. In order to be sure results and posteriors are accurate, fits were run with a high number of live points (500), removing 50 points per run, with 100 inner steps.
Apart from fitting just the foreground to compare constrained and unconstrained fits, we later inject the signal into the data and attempt to retrieve it while cojointly fitting the foreground with the signal:
\begin{equation}
    T_{\mathrm{model}}(\nu) = T_{\mathrm{fg}}(\nu,\boldsymbol{\theta}) + \frac{T_{21}(\nu, \boldsymbol{\phi})}{B_{\mathrm{factor}}(\nu)}.
\end{equation}
Here, $T_{\mathrm{fg}}(\nu,\boldsymbol{\theta})$ is the foreground model  parametrised by coefficients $\boldsymbol{\theta}$, and $T_{21}(\nu,\boldsymbol{\phi})$  is the signal model with parameters $\boldsymbol{\phi} = \{A, \nu_0, w, \tau\}$ for a flattened Gaussian EDGES signal, or $\boldsymbol{\phi} = \{A, \nu_0, w\}$ for the shallower Gaussian.

\begin{table}
\centering
\begin{tabular}{c c c}
\hline
Parameter & Prior & Scale \\
\hline
$\theta_0$ & $\mathcal{N}(c_0,\ 500)$ & fixed \\
$\theta_1$ & $\mathcal{N}(c_1,\ 2|c_1|)$ & $2\times$ polyfit value \\
$\theta_2, \theta_3, \theta_4$ & $\mathcal{N}(c_i,\ |c_i|)$ & $1\times$ polyfit value \\
$\theta_5, \ldots, \theta_{n-1}$ & $\mathcal{N}(c_i,\ 0.1|c_i|)$ & $0.1\times$ polyfit value \\
\hline
$\nu_0$ & $\mathcal{U}(40,\ 120)$ & MHz \\
$w$ & $\mathcal{U}(10,\ 30)$ & MHz \\
$A$ & $\mathcal{U}(0,\ 1)$ & K \\
\hline
$\sigma_\mathrm{noise}$ & $\log\mathcal{N}(\log 0.05,\ 0.1)$ & K \\
\hline
\end{tabular}
\caption{Prior distributions used in the nested sampling analysis for the case where the model is a difference polynomial. 
Foreground polynomial coefficients $\theta_i$ are assigned Gaussian priors centred on 
the corresponding polyfit coefficients $c_i$ obtained by fitting a n=15 polynomial 
to the data, with widths that tighten for higher order terms to regularise the fit. 
Signal parameters are the center frequency ($v_0$), width $w$, and amplitude $A$ of the 
injected Gaussian. The noise standard deviation $\sigma_\mathrm{noise}$ is given a 
log-normal prior (preferred value is 0.05 and it has 0.1 sigma spread in log space).}
\label{tab:diffpoly_priors}
\end{table}

\begin{table}
\centering
\begin{tabular}{c c c}
\hline
Parameter & Prior & Scale \\
\hline
$\theta_0$ & $\mathcal{N}(0,\ 500)$ & fixed \\
$\theta_1, \ldots, \theta_{n-1}$ & $\mathcal{N}(0,\ 100)$ & fixed \\
\hline
$\sigma_\mathrm{noise}$ & $\log\mathcal{N}(\log 0.05,\ 0.1)$ & K \\
\hline
\end{tabular}
\caption{Prior distributions when using a n=5 log-log polynomial model for foreground.}
\label{tab:loglog_priors}
\end{table}

\begin{table}
\centering
\begin{tabular}{c c c}
\hline
Parameter & Prior & Scale \\
\hline
$\theta_0$ & $\mathcal{N}(2300,\ 500)$ & fixed \\
$\theta_1, \ldots, \theta_{n-1}$ & $\mathcal{U}(-15,\ 15)$ & fixed \\
\hline
$\sigma_\mathrm{noise}$ & $\log\mathcal{N}(\log 0.05,\ 0.1)$ & K \\
\hline
\end{tabular}
\caption{Prior distributions when using the n=5 EDGES power law expansion polynomial model for foreground.}
\label{tab:edges_priors}
\end{table}

\subsection{Likelihood and priors}
Given a foreground model $T(\nu,\boldsymbol{\theta})$, the likelihood is assumed to be Gaussian,
\begin{equation}
\label{eqn:likelihood}
\mathcal{L}(\boldsymbol{\theta}) = \frac{1}{(2\pi\sigma)^{N/2}}\exp\!\left(-\frac{1}{2}\sum_{i}^{N} \frac{\left[T_{\mathrm{data}}(\nu_i) - T(\nu_i,\boldsymbol{\theta})\right]^2}{\sigma^2}\right),
\end{equation}
where $\sigma$ is the noise level of the data. Rather than fixing $\sigma$ to an assumed value, we treat it as a free parameter in the fit. This allows the model to self consistently infer the noise level from the residuals, avoiding the risk of over or under constraining the likelihood through an incorrect noise assumption. A broad log uniform prior is placed on $\sigma$ to reflect prior ignorance of its magnitude while ensuring scale invariance.

Priors on the foreground model coefficients are assigned based on physically motivated ranges and numerical stability considerations. For polynomial coefficients, uniform priors are adopted over ranges that fit with the expected order of magnitude of each term given the spectral behaviour of the galactic foreground at the frequencies under consideration. The prior boundaries are set conservatively wide to ensure they do not artificially truncate the posterior, while remaining narrow enough to avoid wasting computational resources exploring regions of negligible posterior mass.

The same three foreground parameterisations used in the constrained fits are adopted here: the log-log polynomial, the difference polynomial, and the EDGES style power law expansion. In each case no derivative constraints are imposed, and all model parameters are free to vary continuously within their prior distributions, which are given in Tables \ref{tab:diffpoly_priors} for a difference polynomial, \ref{tab:loglog_priors} for the log-log polynomial and \ref{tab:edges_priors} for the EDGES power law expansion. The number of terms in each parameterisation is varied as in the constrained analysis, allowing us to assess the evidence for each model order and compare across parameterisations using Bayes factors.

\begin{figure}
    \centering
    \includegraphics[width=\columnwidth]{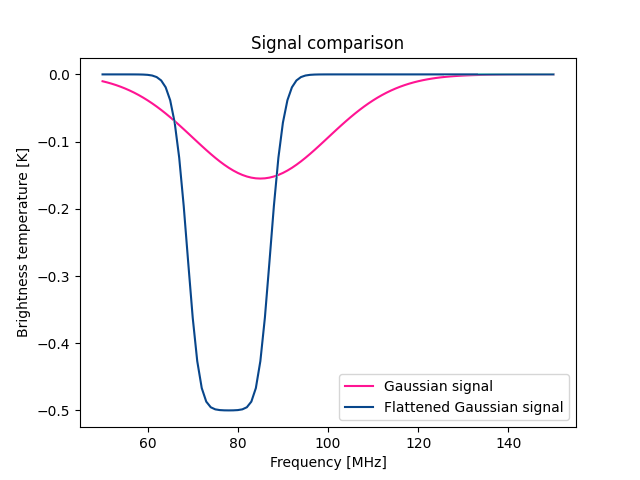}
    \caption{The injected signals used to test the signal recovery under different foreground maps and chromaticity corrections.}
    \label{fig:injectedsignals}
\end{figure}

\subsection{Comparison with constrained fitting}
\label{sec:comparison}

Unlike \textsc{maxsmooth}, the unconstrained fits impose no derivative-sign restriction, so the foreground model is free to absorb genuine signal, systematics, or noise
at high polynomial order. Comparing the two frameworks therefore lets us distinguish features that are robust to the smoothness prior, and so have a
stronger claim to being real, from those that are sensitive to it. The unconstrained framework additionally yields full posteriors, rather than \textsc{maxsmooth}'s point estimates, allowing uncertainty on the recovered signal and foreground parameters to be quantified directly. In principle, the Bayesian evidence computed by nested sampling could also be used to formally select the
polynomial order for each parameterisation; in this work we instead choose the order for each model based on the point at which further terms cease to reduce
the residuals (\S\ref{sec:results}).

\begin{figure*}
    \centering

    \begin{subfigure}{0.49\textwidth}
        \centering
        \includegraphics[width=\linewidth]{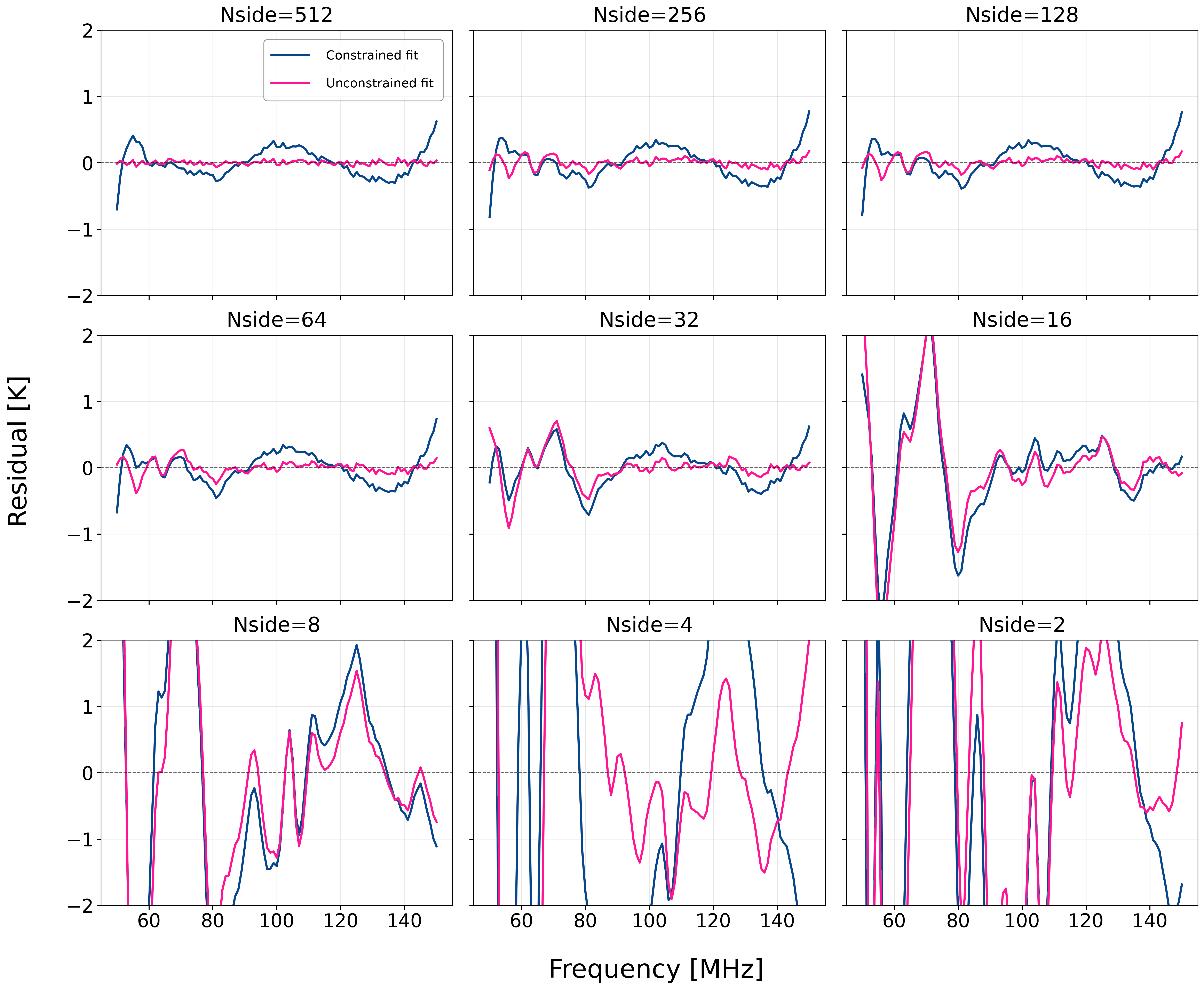}
        \caption{Uniform spectral index }
        \label{fig:uniform_edges}
    \end{subfigure}
    \hfill
    \begin{subfigure}{0.49\textwidth}
        \centering
        \includegraphics[width=\linewidth]{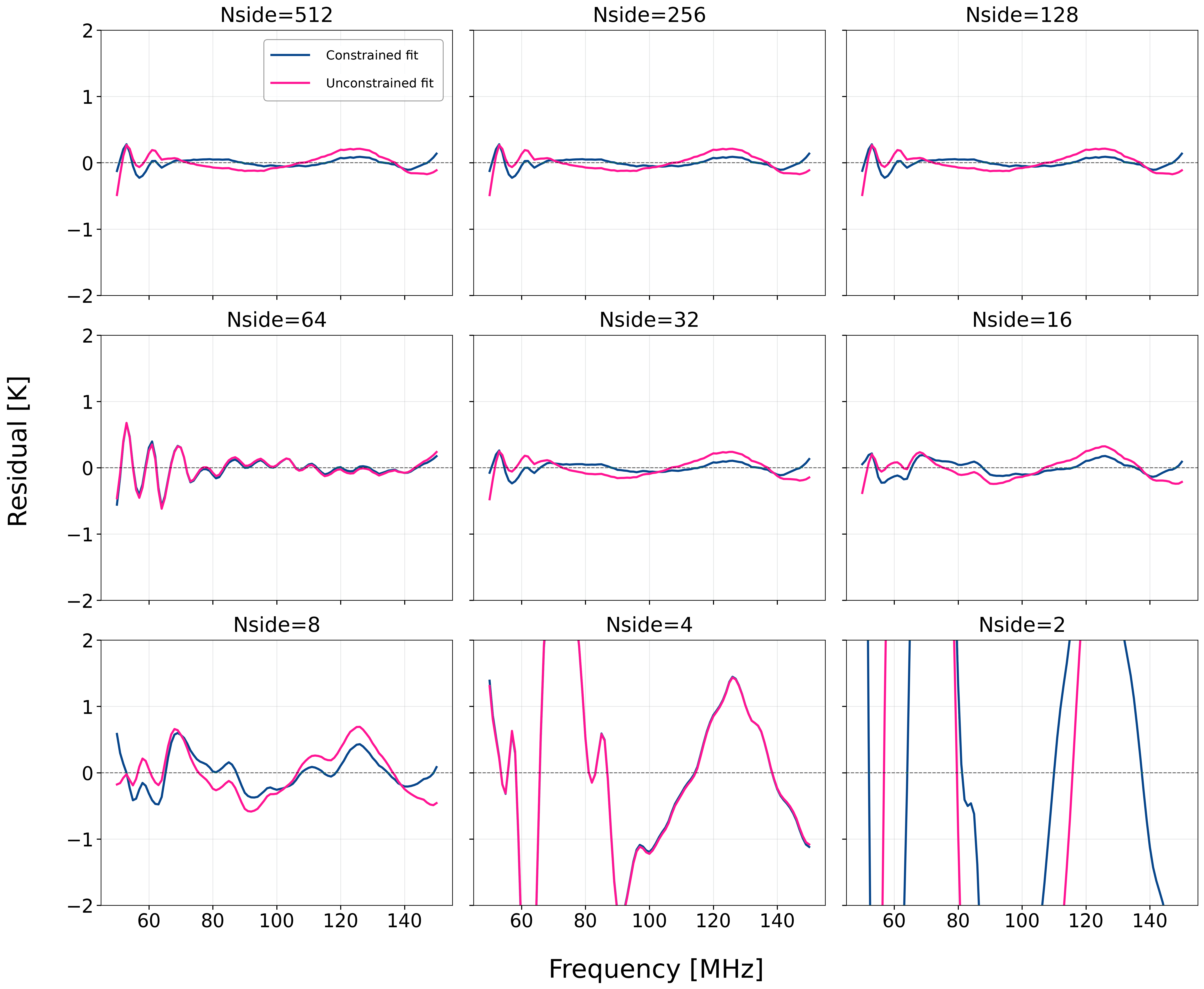}
        \caption{Spatially varying spectral index }
        \label{fig:full_sp_edges}
    \end{subfigure}

    \caption{Residuals after applying chromaticity corrections to data assuming a uniform spectral index (left) and a spatially varying spectral index (right). The data are fitted using constrained (\textsc{maxsmooth}) and unconstrained foreground models for the EDGES-style power-law expansion parameterisation.}
    \label{fig:edges_fit}
\end{figure*}

\begin{figure*}
    \centering

    \begin{subfigure}{0.49\textwidth}
        \centering
        \includegraphics[width=\linewidth]{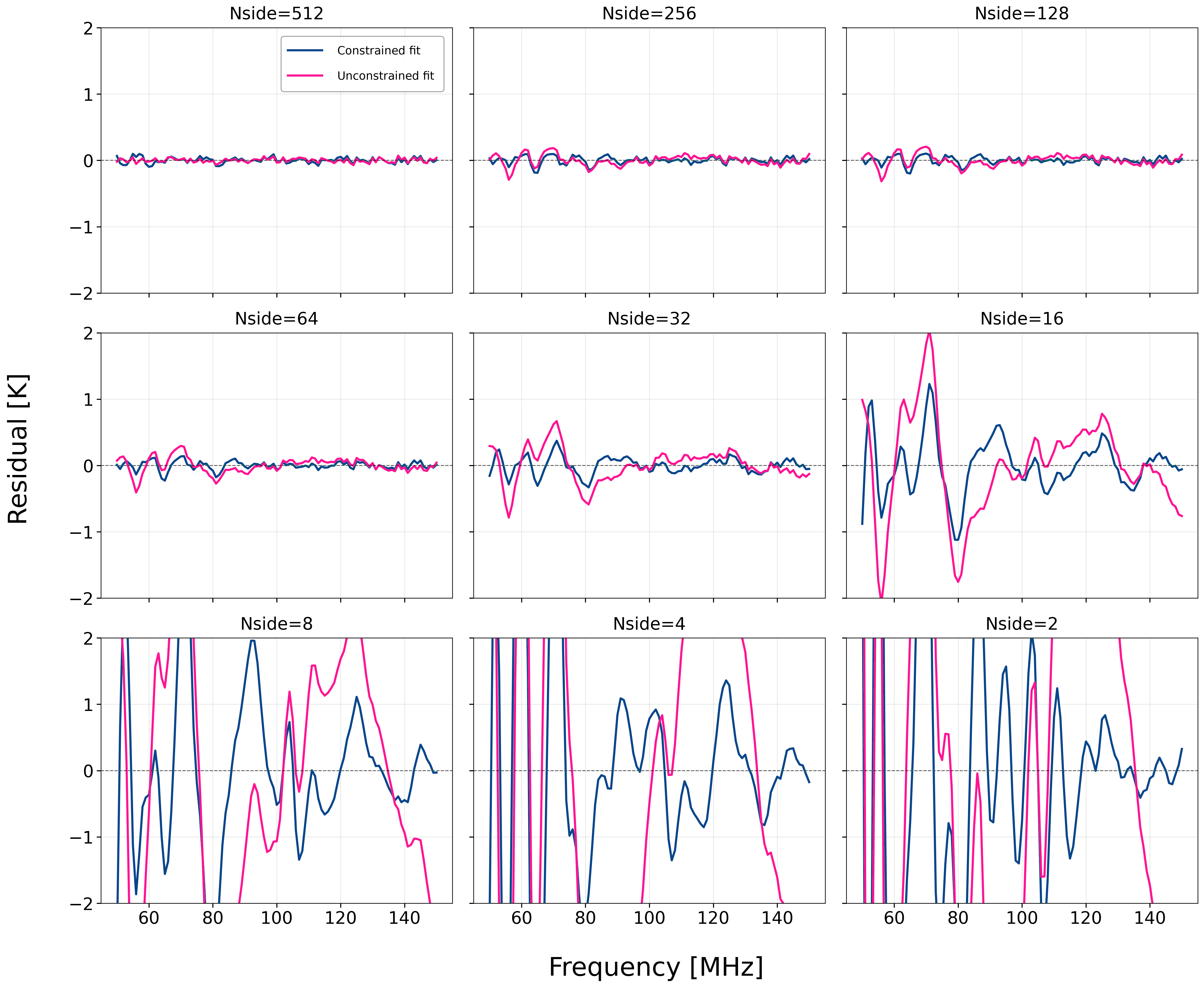}
        \caption{Uniform spectral index}
        \label{fig:uniform_loglog}
    \end{subfigure}
    \hfill
    \begin{subfigure}{0.49\textwidth}
        \centering
        \includegraphics[width=\linewidth]{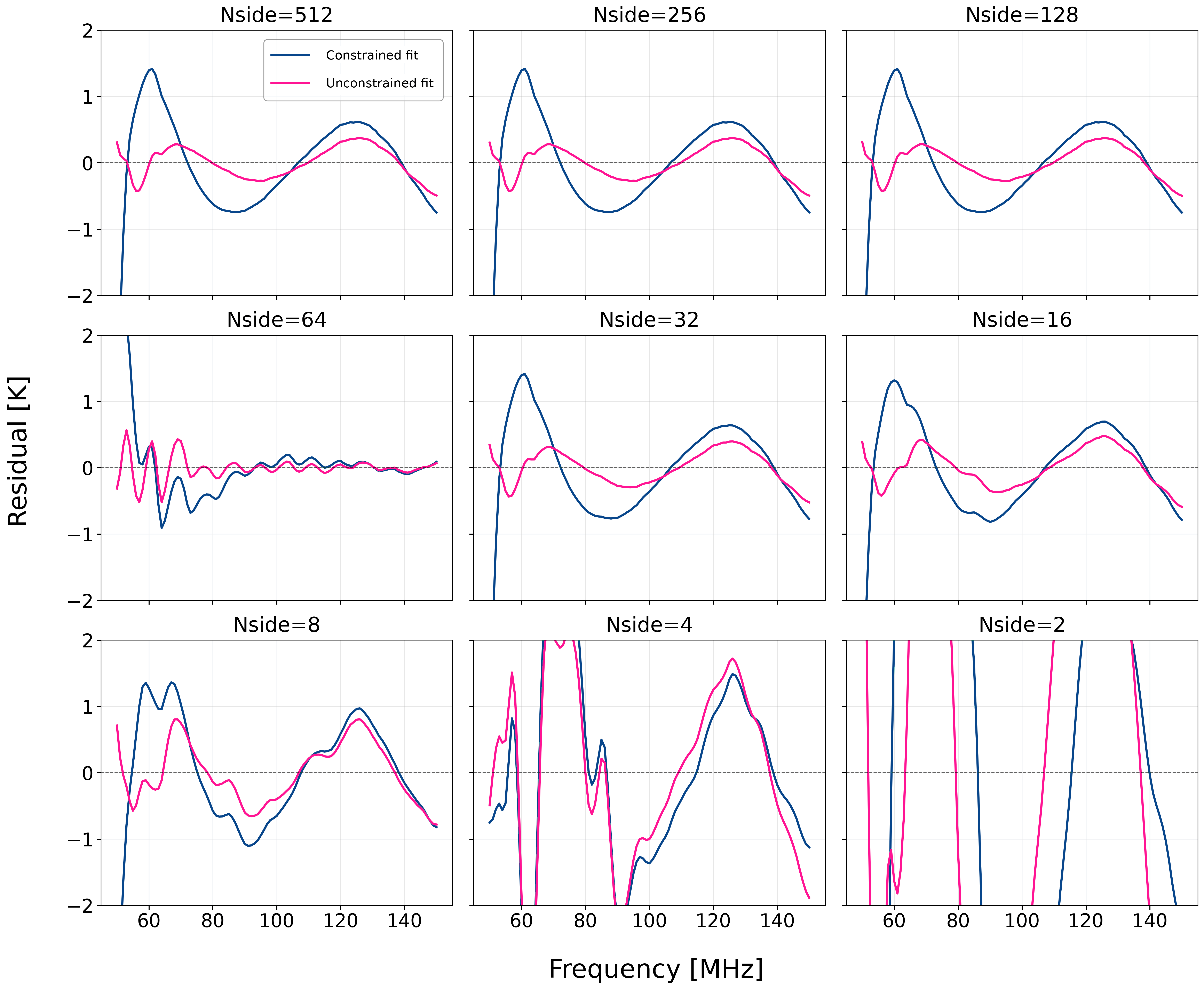}
        \caption{Spatially varying spectral index}
        \label{fig:full_sp_loglog}
    \end{subfigure}

    \caption{Residuals after applying chromaticity corrections to data assuming a spatially varying spectral index (right) and uniform (left). The data are fitted using constrained (\textsc{maxsmooth}) and unconstrained foreground models for the log-log polynomial parameterisation.}
    \label{fig:loglog_fit}
\end{figure*}

\begin{figure*}
    \centering

    \begin{subfigure}{0.49\textwidth}
        \centering
        \includegraphics[width=\linewidth]{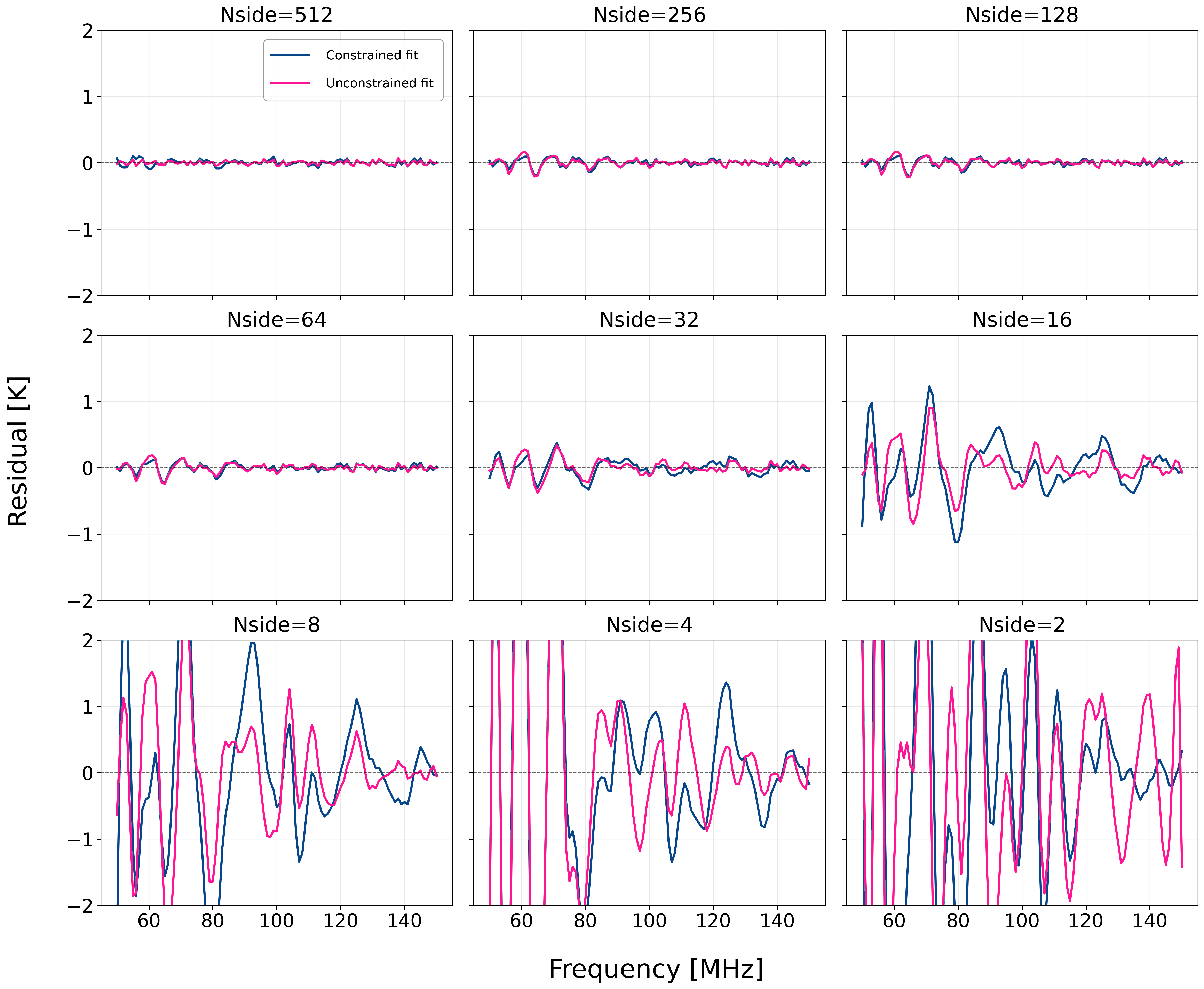}
        \caption{Uniform spectral index}
        \label{fig:uniform_diff}
    \end{subfigure}
    \hfill
    \begin{subfigure}{0.49\textwidth}
        \centering
        \includegraphics[width=\linewidth]{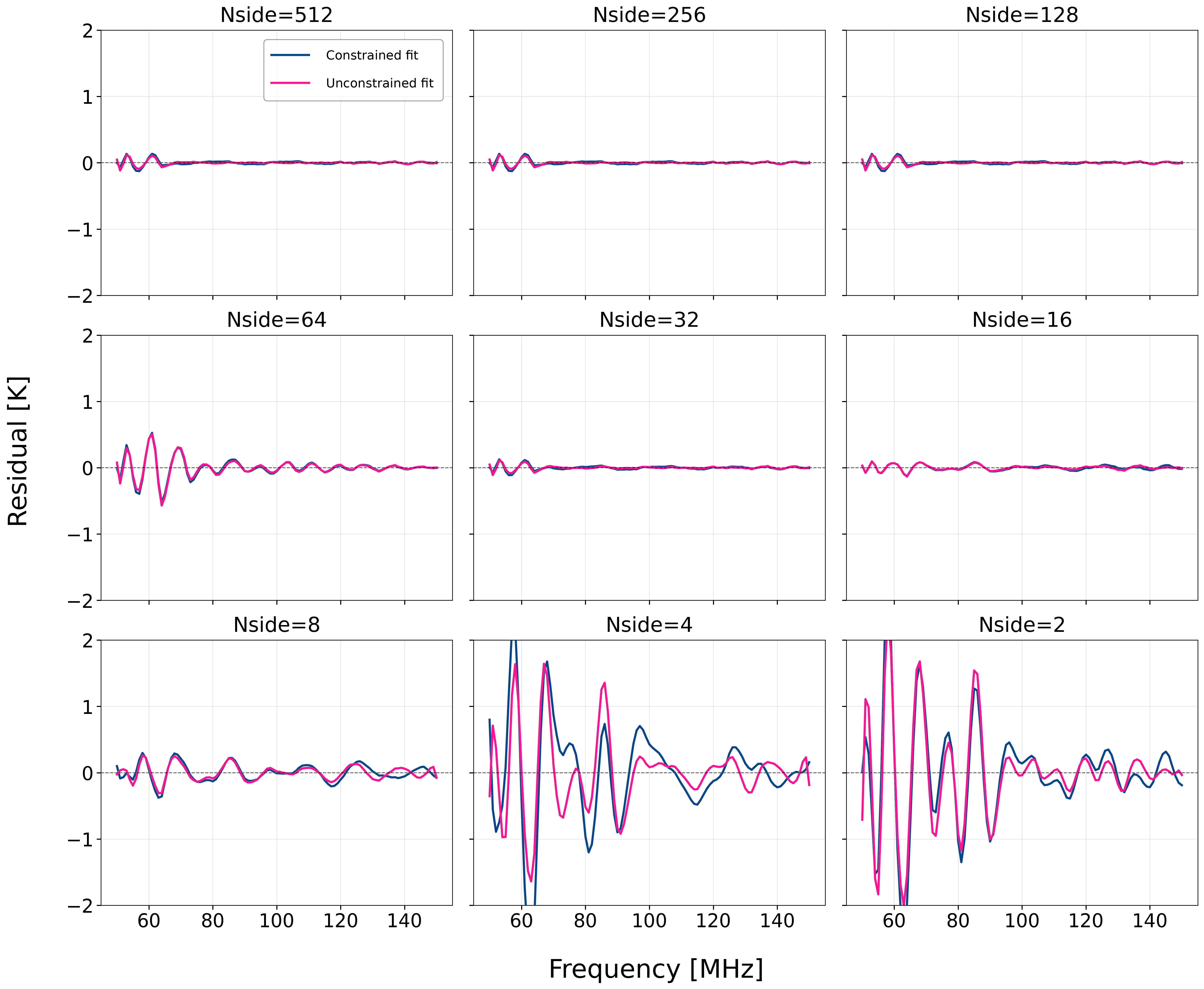}
        \caption{Spatially varying spectral index}
        \label{fig:full_sp_diff}
    \end{subfigure}

    \caption{Residuals after applying chromaticity corrections to data assuming a spatially varying spectral index (right) and uniform (left). The data are fitted using constrained (\textsc{maxsmooth}) and unconstrained foreground models for the difference polynomial parameterisation.}
    \label{fig:diff_fit}
\end{figure*}

\section{Results}
\label{sec:results}

\begin{figure}
    \centering
    \includegraphics[width=\columnwidth]{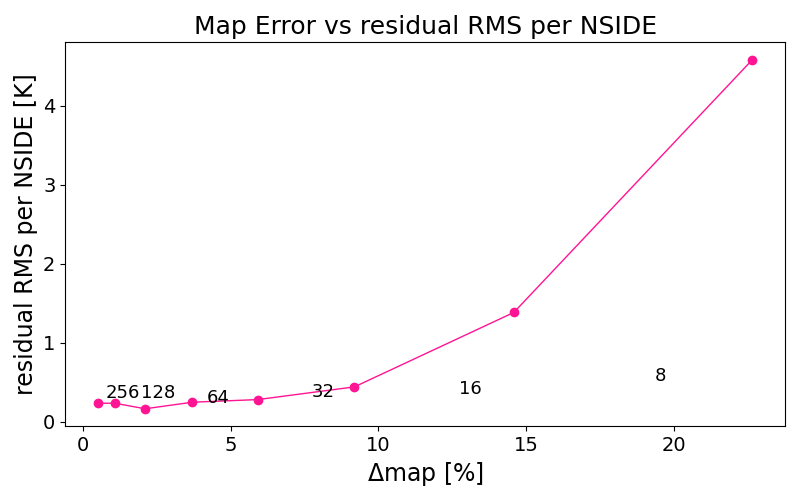}
    \caption{For a single data set, here we show how the residuals after correcting for chromaticity and subtracting the foreground are affected by the error introduced into the sky map by blurring it. Each point represents the RMS of the residuals, $\mathrm{RMS}(\Delta T) = \sqrt{\frac{1}{N_\nu}\sum_{i=1}^{N_\nu} \Delta T_i^2}$, for a given $N_{\rm side}$. Foreground used here is the previously described log-log polynomial fitted without any constraints. RMS($\Delta T$)} is shown against pixel-averaged map error in percentage (see Eq. \ref{eqn:diffpercent}).
    \label{fig:error_vs_residuals}
\end{figure}

Our central result is that residuals after subtracting the foreground and correcting for chromaticity are insensitive to sky-map resolution down to $N_{\rm side} = 64$ (map error of 2.1\% which translates to an angular resolution of $\sim 1^\circ$) when the data and correction assume uniform spectral index, regardless of which foreground model was used, or which fitting framework was used (Figures.~\ref{fig:uniform_edges}, \ref{fig:uniform_loglog}, \ref{fig:uniform_diff}). We see a similar trend with our more realistic case that assumes a spatially varying spectral index (Figures.~\ref{fig:full_sp_edges}, \ref{fig:full_sp_loglog}, \ref{fig:full_sp_diff}), but we can also see that residuals at lower resolutions of the map used in the correction like $N_{\rm side} = 32$, $N_{\rm side} = 16$, stay low. This is most likely a result of the sky-map features smoothing out and being suppressed with resolution degradation. Smoother systematics are easier for any spectrally smooth foreground model to absorb, and this combined with the choice of foreground models themselves leads to the residuals looking small. We quantify the goodness of a fit in these figures as the RMS value in Kelvin. RMS increases with the introduced map error in the chromaticity correction and we visualise this in Figure \ref{fig:error_vs_residuals}.
We now examine each foreground parameterisation in turn, followed by the implications for signal recovery.

\subsection{Foreground modelling on degraded maps}

Figures~\ref{fig:edges_fit}, \ref{fig:loglog_fit} and \ref{fig:diff_fit} summarize how fitting residuals degenerate with map degradation in chromaticity correction. Each figure shows both cases described in Section \ref{sec:chromaticity}, assuming $\beta=-2.5$ on the left, and spatially varying $\beta$ on the right.
In Figure~\ref{fig:edges_fit}, the foreground was modelled by five term constrained and unconstrained polynomials described in Eq. \ref{eqn:edges_powerlaw_intro}. 3-term and 7-term versions were also tested. The power-law basis is physically motivated by the $\nu^{-2.5}$ spectral behaviour of Galactic synchrotron emission, so the foreground is already close to a single term in this basis. As a result, a low-order expansion captures most of the spectral structure with minimal truncation error, and the constrained and unconstrained fits perform comparably at the same polynomial order. Residuals remain effectively unchanged down to 2.1\% map divergence from the original.

The full resolution at $N_{\rm side} = 512$ in the case where data generation and chromaticity correction generation assume uniform spectral index should in principle give residuals that contain only noise, and with all three models this is the case. Any discrepancies between residuals is solely due to different fitting methods and different model selection, and these different artifacts introduced into the residuals coming from the fitting models themselves are more prominent in the case where data assumes spatially varying spectral index.

In contrast to using the EDGES power-law expansion to model the foreground, difference polynomial in Eq.~\ref{eqn:differencepolynomial_intro} and log-log polynomial in Eq.~\ref{eqn:loglog_polynomial_intro} are not naturally aligned with the power-law foreground shape. Fitting these basis functions to a foreground that follows a power-law is analogous to truncating a Taylor expansion: an unconstrained fit can redistribute it's coefficients to absorb the truncation error and still achieve a good fit at low order, but once smoothness constraints are imposed the model loses that flexibility and many more terms are required to represent the data equally well. Therefore, \textsc{maxsmooth} requires higher order polynomials to achieve comparable residuals.

Figures~\ref{fig:loglog_fit} and \ref{fig:diff_fit} show residuals after subtracting foreground modeled as a difference polynomial and log-log polynomial respectively. Constrained fits use $N=15$ terms in both cases. Unconstrained log-log foreground model uses $N=5$ terms to get comparable residuals, and the difference polynomial model uses $N=12$ terms, and although most higher terms in this case are very small, using lower order polynomials resulted in 100 K amplitude residuals with ripple-like structures originating from frequency dependent effects from the instrument. These ripples are also noticeable when fitting a constrained log-log polynomial on data that assumes spatially varying spectral index map. These structures aren't spectrally smooth so constraining derivatives doesn't help to absorb them. All residuals consistently start degenerating at $N_{\rm side} = 32$, with the amplitude exploding at $N_{\rm side} = 16$ where the map error is sufficiently large for the chromaticity correction to fail. This trend is visually represented in Figure \ref{fig:error_vs_residuals}. Residual RMS is staying roughly constant, and starts growing at $N_{\rm side} = 16$ where map error is $\sim 6\%$ $\sim 3[K]$. It exponentially increases by 3 Kelvin as the map is fully blurred.

\begin{figure*}
    \centering

    \begin{subfigure}{0.49\textwidth}
        \centering
        \includegraphics[width=\linewidth]{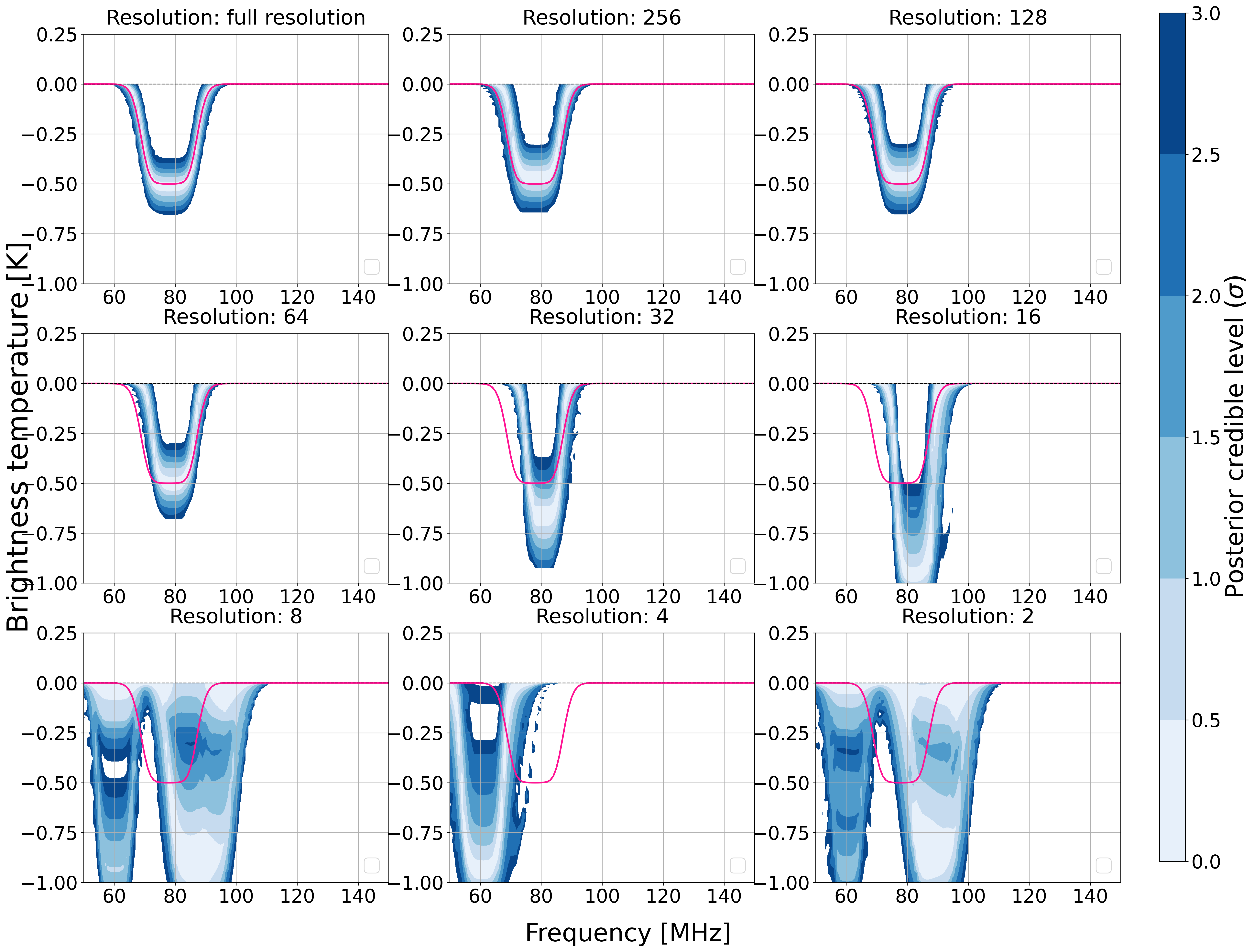}
        \caption{Uniform spectral index (EDGES power-law expansion)}
        \label{fig:uniform_edges_fgauss}
    \end{subfigure}
    \hfill
    \begin{subfigure}{0.49\textwidth}
        \centering
        \includegraphics[width=\linewidth]{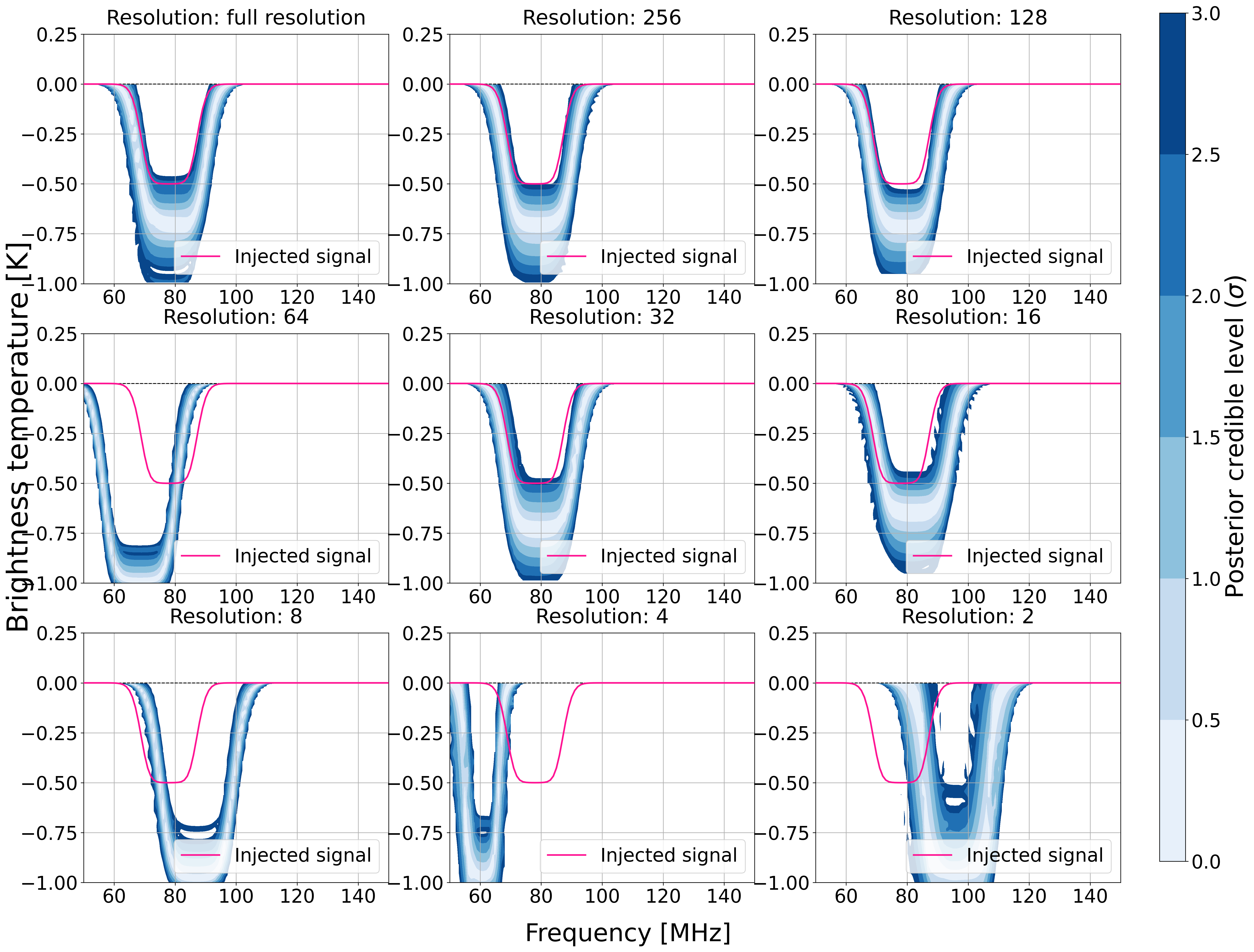}
        \caption{Spatially varying spectral index (EDGES power-law expansion)}
        \label{fig:full_sp_edges_fgauss}
    \end{subfigure}

    \caption{Recovered signal posteriors in comparison to the injected flattened gaussian. In this case data was fit with the EDGES power law expansion described in Eq. \ref{eqn:edges_powerlaw_intro}}
    \label{fig:edges_fgauss}
\end{figure*}

\begin{figure*}
    \centering

    \begin{subfigure}{0.49\textwidth}
        \centering
        \includegraphics[width=\linewidth]{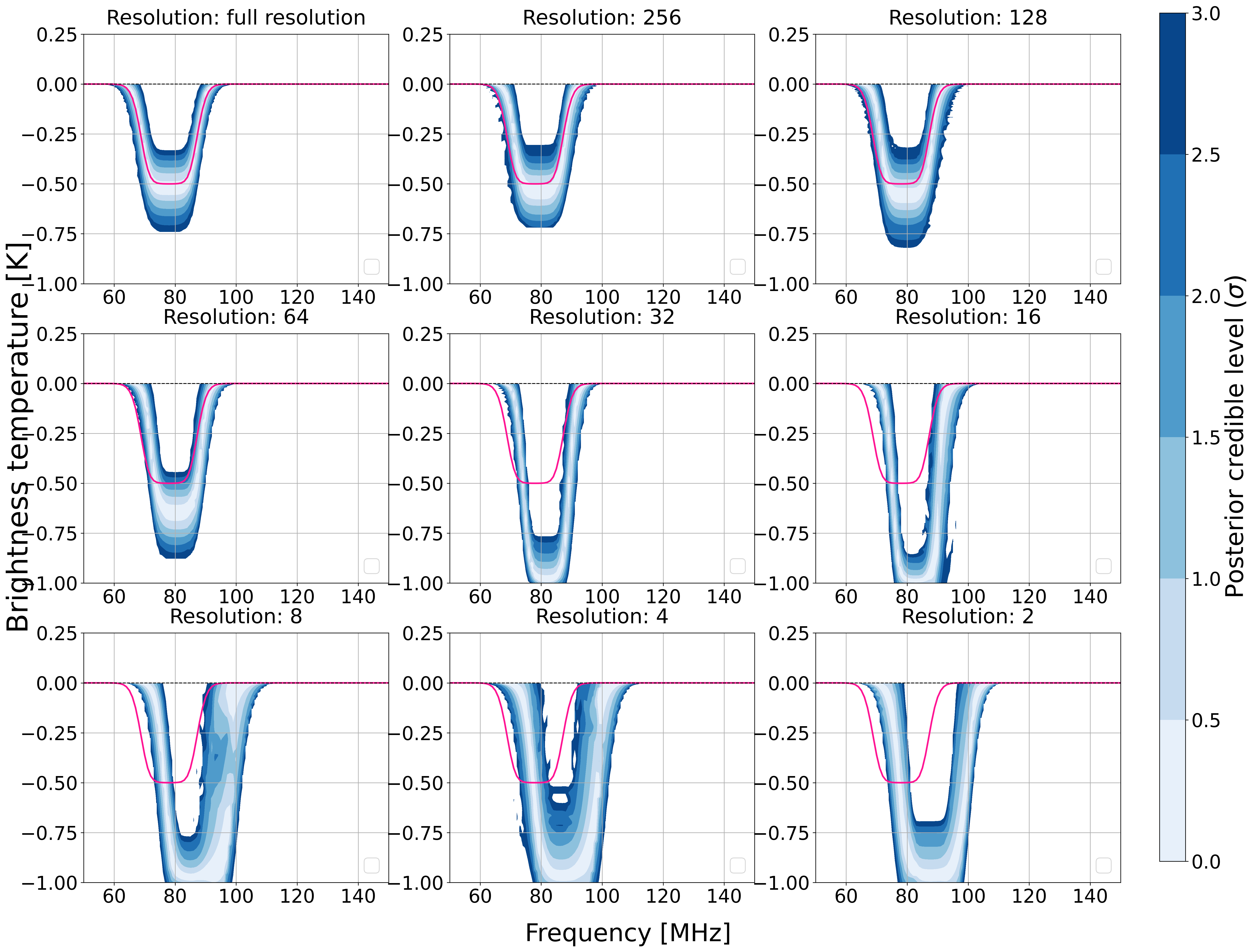}
        \caption{Uniform spectral index}
        \label{fig:uniform_loglog_fgauss}
    \end{subfigure}
    \hfill
    \begin{subfigure}{0.49\textwidth}
        \centering
        \includegraphics[width=\linewidth]{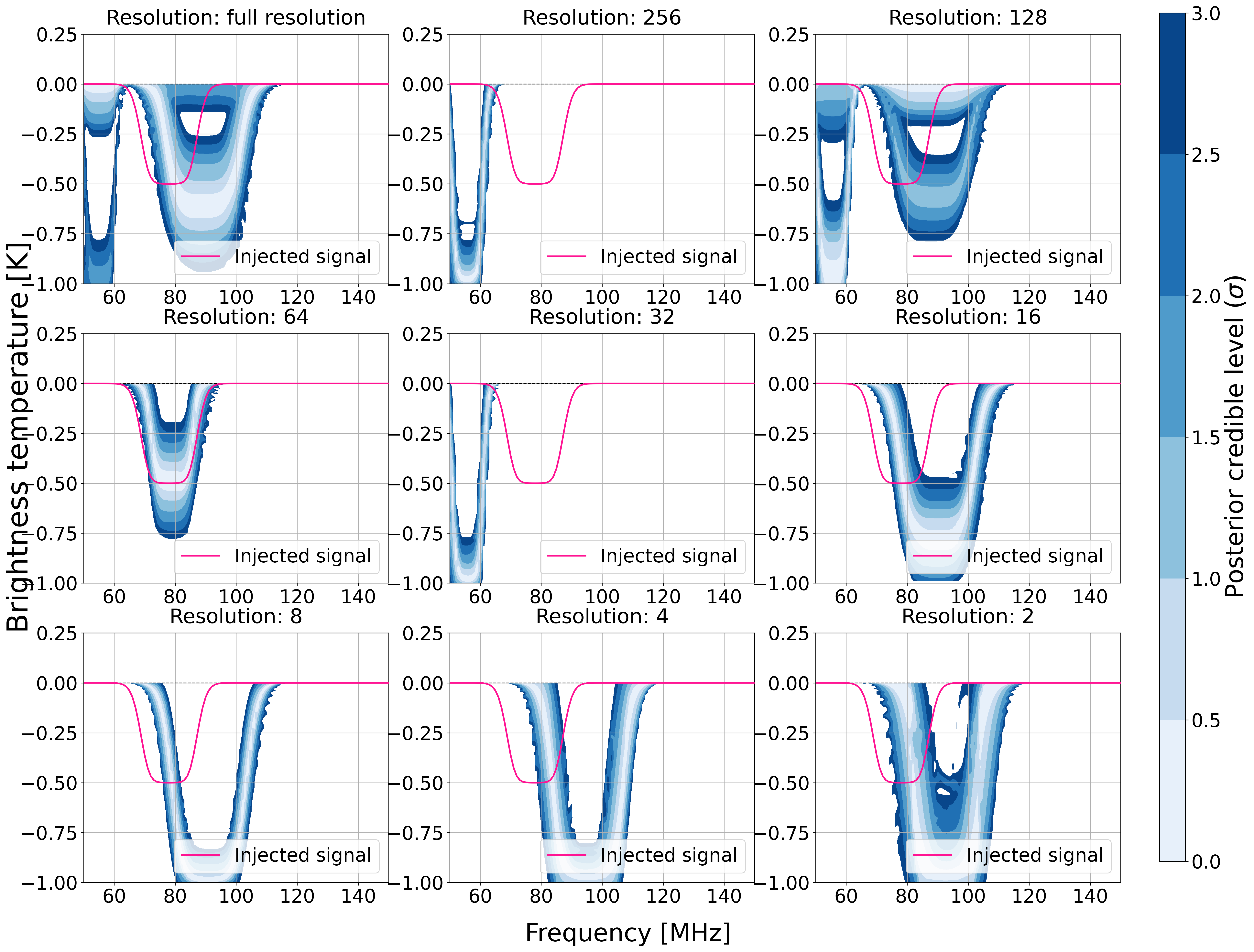}
        \caption{Spatially varying spectral index}
        \label{fig:full_sp_loglog_fgauss}
    \end{subfigure}

    \caption{Recovered signal posteriors in comparison to the injected flattened gaussian. In this case data was fit with the log-log polynomial described in Eq. \ref{eqn:loglog_polynomial_intro}}
    \label{fig:loglog_fit_fgauss}
\end{figure*}

\begin{figure*}
    \centering

    \begin{subfigure}{0.49\textwidth}
        \centering
        \includegraphics[width=\linewidth]{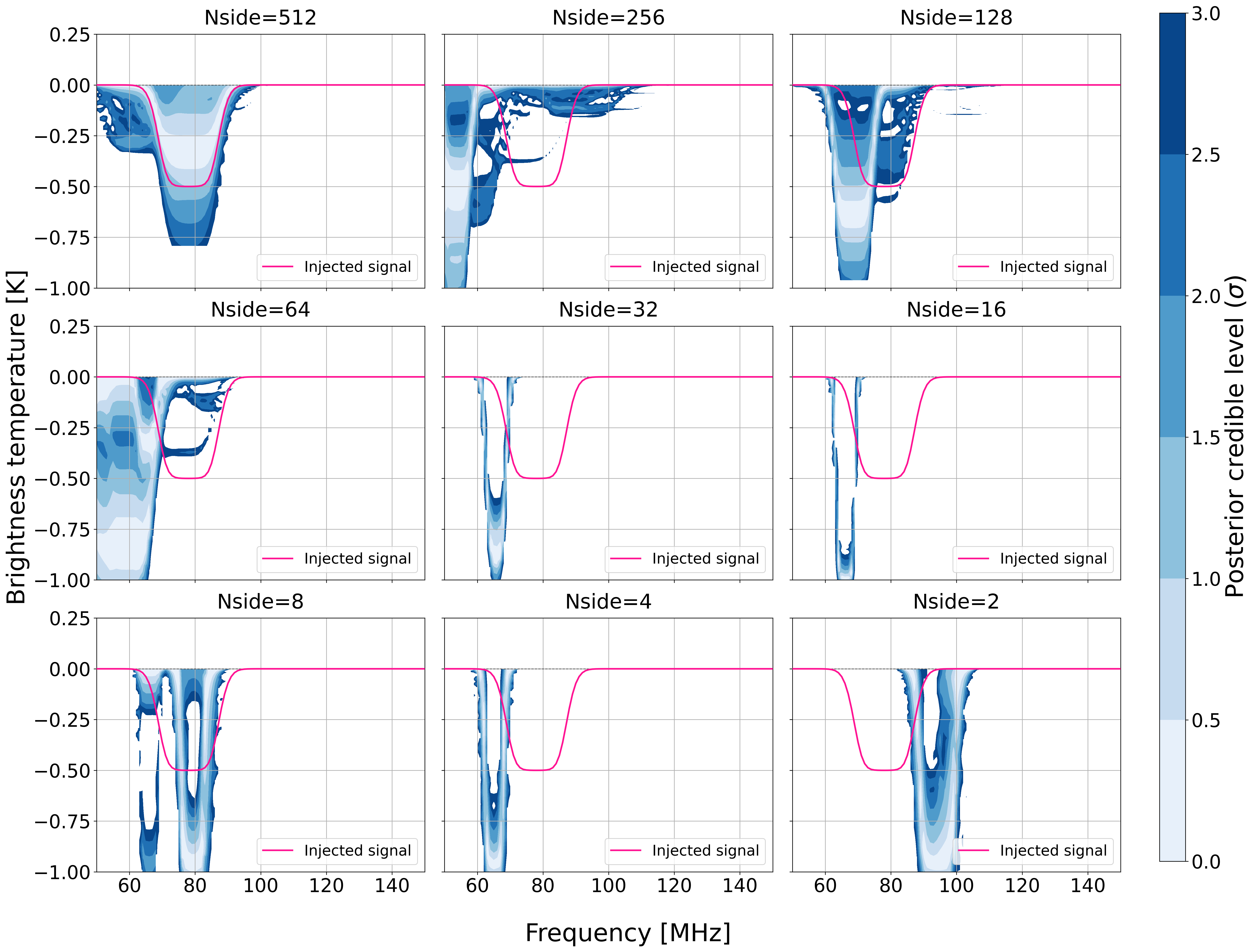}
        \caption{Uniform spectral index}
        \label{fig:uniform_diff_fgauss}
    \end{subfigure}
    \hfill
    \begin{subfigure}{0.49\textwidth}
        \centering
        \includegraphics[width=\linewidth]{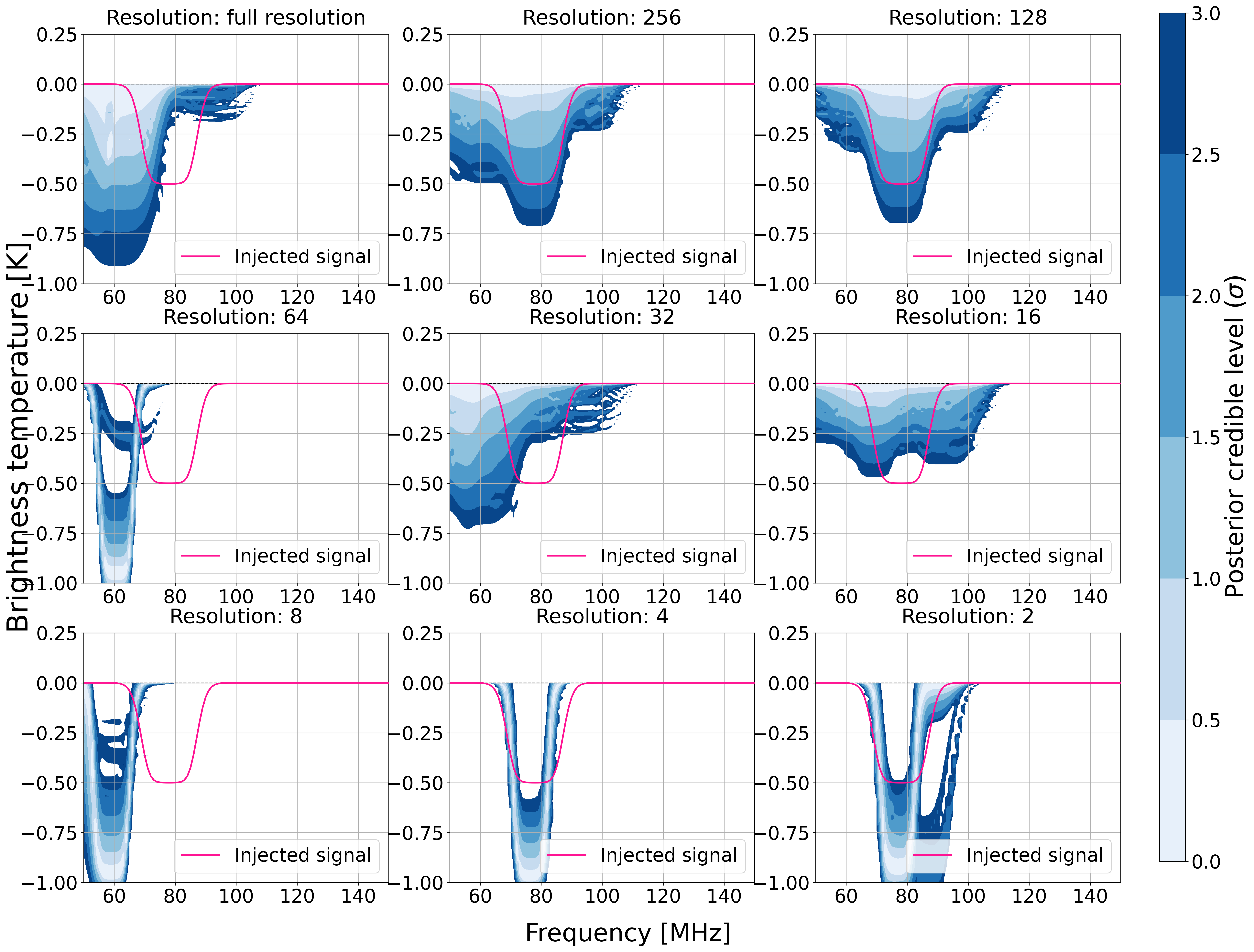}
        \caption{Spatially varying spectral index}
        \label{fig:full_sp_diff_fgauss}
    \end{subfigure}

    \caption{Recovered signal posteriors in comparison to the injected flattened gaussian. In this case data was fit with the difference polynomial described in Eq. \ref{eqn:differencepolynomial_intro}}
    \label{fig:diff_fit_fgauss}
\end{figure*}

\subsection{Signal extraction}
\subsubsection{Flattened Gaussian}

We tested signal recovery using a flattened Gaussian signal described in \cite{EDGES2017} as:
\begin{equation}
T_{21}(\nu) = -A \left( \frac{1 - e^{-\tau e^{B}}}{1 - e^{-\tau}} \right)
\end{equation}
where
\begin{equation}
B = \frac{4(\nu - \nu_{0})^{2}}{w^{2}} 
\ln \left[ -\frac{1}{\tau} \ln \left( \frac{1 + e^{-\tau}}{2} \right) \right],
\end{equation}
$\tau = 7$ is the flattening factor, $w=18.7$ MHz is the width parameter, and $A=0.5$ K and $\nu_{r} = 78$ MHz are the amplitude and center frequency respectively. We attempt to retreive the injected signal from the data by cojointly fitting the signal and the three different foregrounds to see how foreground selection impacts signal retrieval.
In Figures \ref{fig:edges_fgauss}, \ref{fig:loglog_fit_fgauss} and \ref{fig:diff_fit_fgauss} we show the results of signal retrieval. The coloured contours follow the \texttt{fgivenx} convention \citep{fgivenx}, where lighter shading corresponds to higher posterior probability density, the lightest region encloses the most probable signal shapes, analogous to a $1\sigma$ credible interval, with darker bands extending to $2\sigma$ and $3\sigma$. The injected flattened Gaussian is overlaid in pink for direct comparison.

In this section we use the unconstrained (Bayesian nested sampling) fits described in \S\ref{sec:unconstrained} to recover the injected flattened Gaussian signal, and examine how recovery quality changes with sky-map resolution. As expected, recovery quality degrades with resolution degradation, consistent with the residual trends reported above. The results for the power-law expansion, log-log polynomial, and difference polynomial are shown in Figures~\ref{fig:edges_fgauss}, \ref{fig:loglog_fit_fgauss}, and \ref{fig:diff_fit_fgauss} respectively. When both the injected data and the chromaticity correction assume a uniform spectral index ($\beta = \mathrm{const.}$), the signal is recovered accurately, with narrow posterior credible intervals; this is the case for all three foreground models, but is cleanest for the EDGES power-law expansion, whose low residuals allow the injected 0.5\,K depth to be recovered reliably. When a spatially varying spectral index is instead used for both the injected
data and the fitted foreground, recovery is markedly worse: even using the same EDGES power-law foreground model, the recovered posterior consistently prefers a deeper absorption trough than the one injected, using the priors described in Table \ref{tab:flattened}.

\begin{table}
\centering
\begin{tabular} {cccc}
\hline
Parameter & Prior & Units \\
\hline
$\nu_0$ & $\mathcal{U}(50,\ 150)$ & MHz \\
$w$ & $\mathcal{U}(5,\ 30)$ & MHz \\
$A$ & $\mathcal{U}(0,\ 1)$ & K \\
$\tau$ & $\mathcal{U}(3,\ 20)$  \\
\hline
\end{tabular}
\caption{Prior distributions for the signal parameters, $\nu_0$ is the central frequency, $w$ width of the flattened bottom, $A$ amplitude and $\tau$ flattening parameter. These priors were used in all cases where a flattened Gaussian was injected as a signal.}
\label{tab:flattened}
\end{table}

Central frequency priors were purposefully left as very wide as seen in Table \ref{tab:flattened} to eliminate/or confirm the hypothesis that chromatic effects are being mistaken as the signal and being picked up by the sampler. Signal retrieval becomes very inaccurate very fast. Posterior credibility ranges are wide, so even though the residuals we see are low and seem to get rid of chromaticity perfectly, there is some level of degeneracy while cojointly fitting data with injected signal, between the amplitude of the signal and the data. 

Using the difference polynomial from Eq.~\ref{eqn:differencepolynomial_intro} to model the foreground gave the lowest residuals but the worst posterior probabilities for signal extraction. This could be due to number of reasons, including that a very high term polynomial is needed to fit the data, so the signal was most likely fit with the foreground. A less flexible polynomial such as the log-log polynomial with 5 terms seems to do much better.

\subsubsection{Gaussian}
\label{gaussian}
A more widely accepted model for the 21-cm Cosmic Dawn signal is a much shallower Gaussian shaped one. We use a Gaussian shaped signal that is not directly physically motivated:
\begin{equation}
T_{21}(\nu) =
-A \exp\left[
-\frac{(\nu-\nu_0)^2}{2w^2}
\right],
\label{eq:gaussian_signal}
\end{equation}
although all parameters are based on physically motivated values: the amplitude is assumed to be A = 0.155 K, $\nu_0 = 85$ MHz central frequency, and FWHM $w = 15$ MHz . Priors used for cases where this signal was injected are shown in Table \ref{tab:diffpoly_priors}.

\begin{figure*}
    \centering

    \begin{subfigure}{0.49\textwidth}
        \centering
        \includegraphics[width=\linewidth]{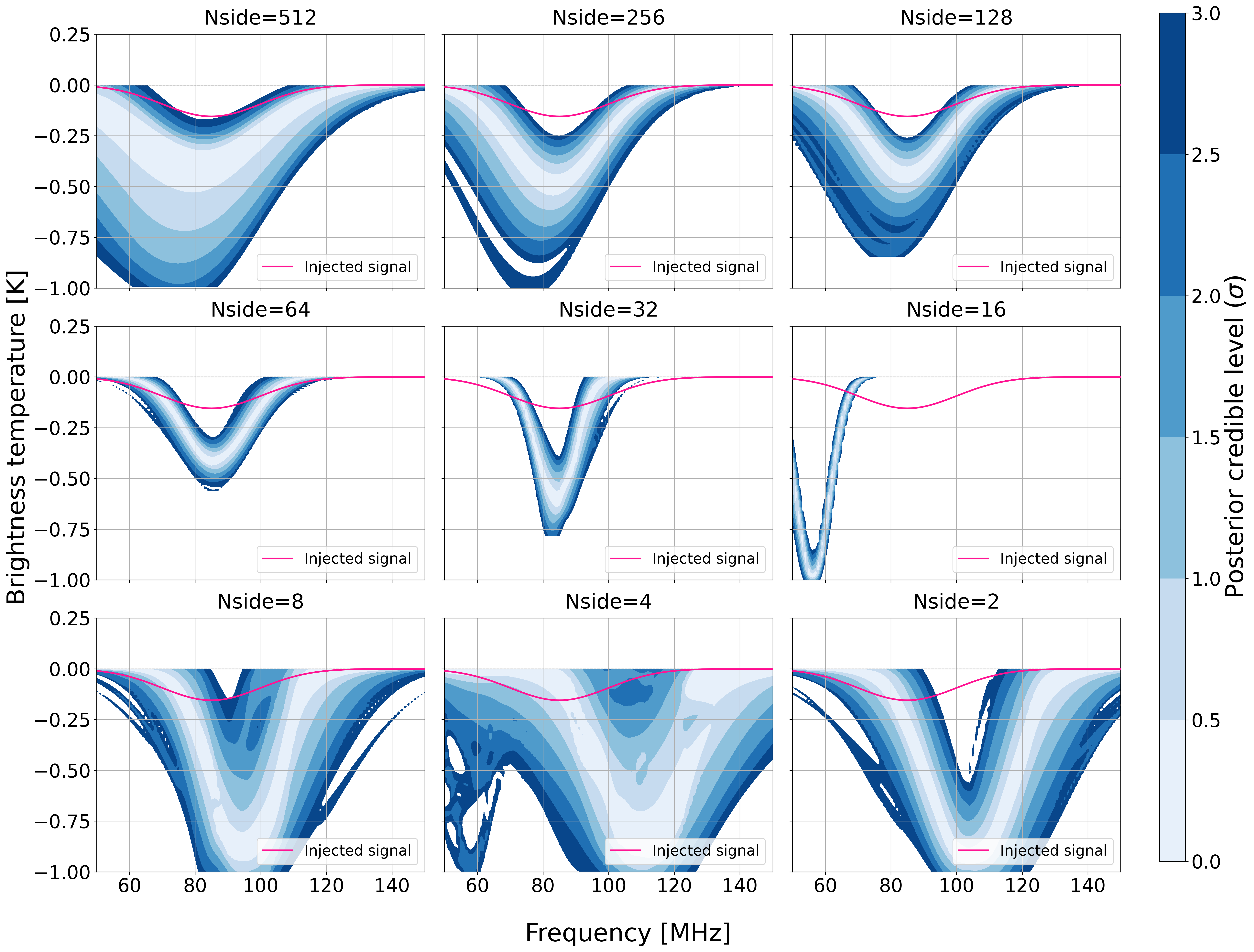}
        \caption{Uniform spectral index (EDGES power-law expansion)}
        \label{fig:uniform_edges_gauss}
    \end{subfigure}
    \hfill
    \begin{subfigure}{0.49\textwidth}
        \centering
        \includegraphics[width=\linewidth]{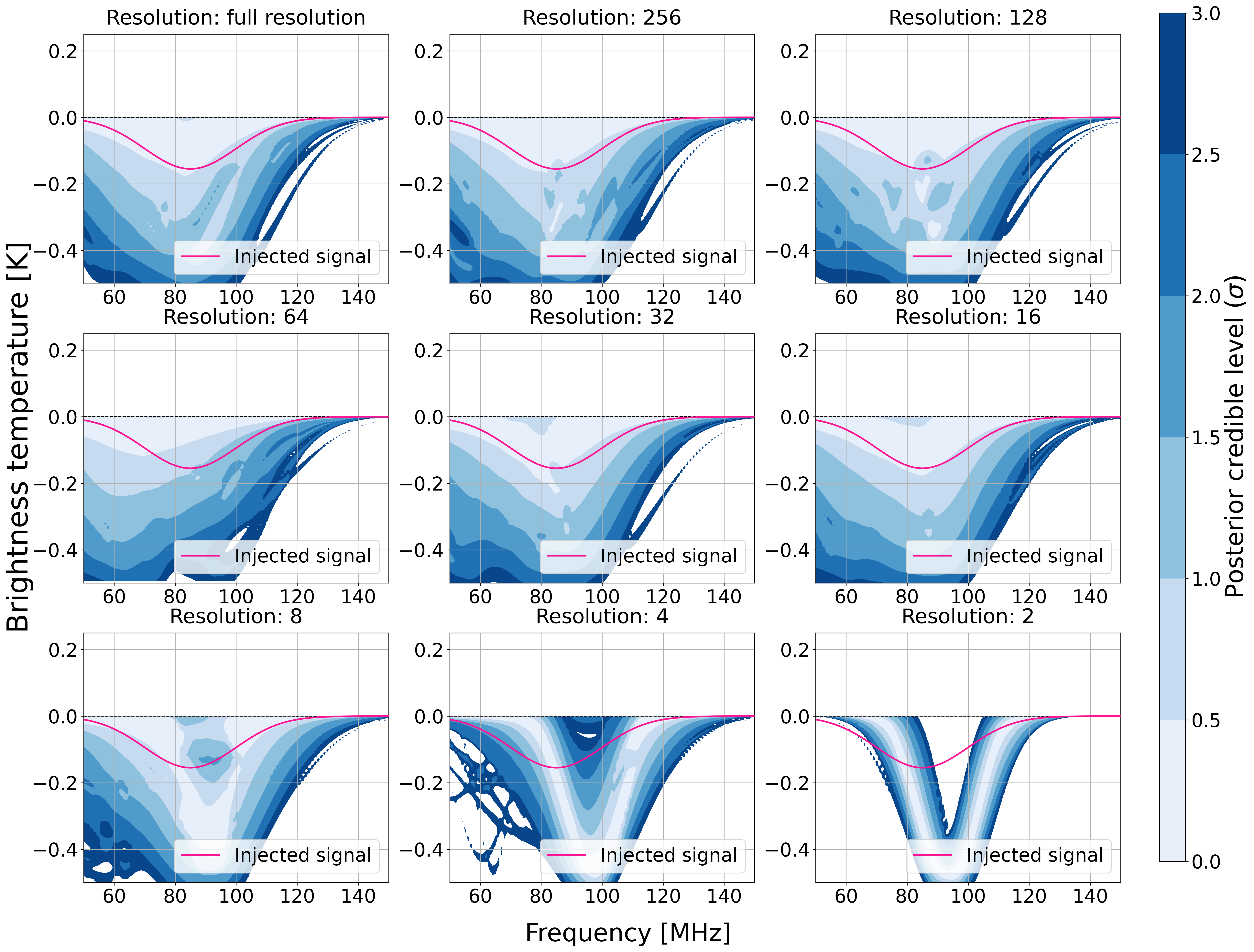}
        \caption{Spatially varying spectral index}
        \label{fig:full_sp_edges_gauss}
    \end{subfigure}

    \caption{Recovered signal posteriors in comparison to the injected Gaussian. In this case data was fit with the EDGES power law expansion described in \ref{eqn:edges_powerlaw_intro}}
    \label{fig:edges_gauss}
\end{figure*}

\begin{figure*}
    \centering

    \begin{subfigure}{0.49\textwidth}
        \centering
        \includegraphics[width=\linewidth]{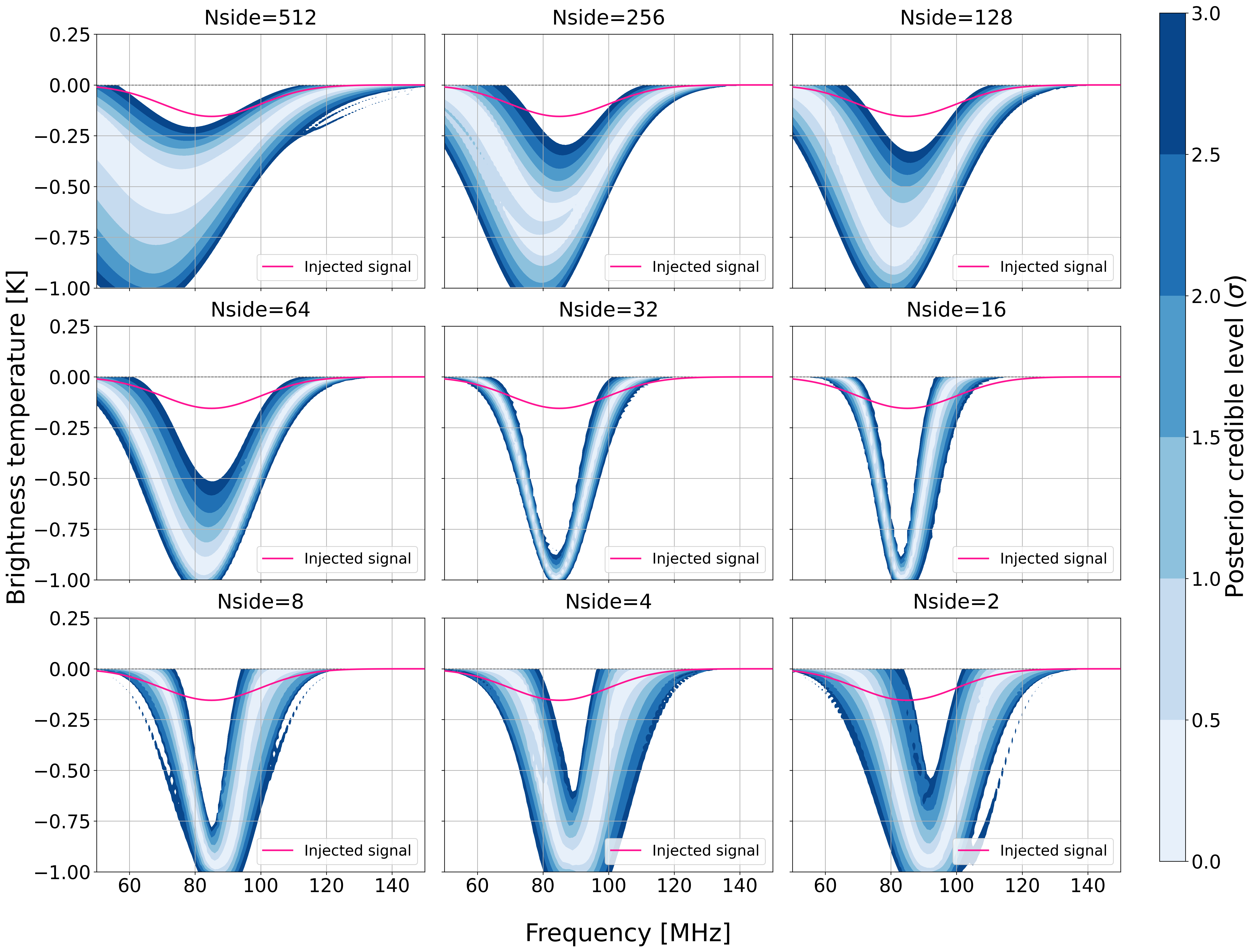}
        \caption{Uniform spectral index}
        \label{fig:uniform_loglog_gauss}
    \end{subfigure}
    \hfill
    \begin{subfigure}{0.49\textwidth}
        \centering
        \includegraphics[width=\linewidth]{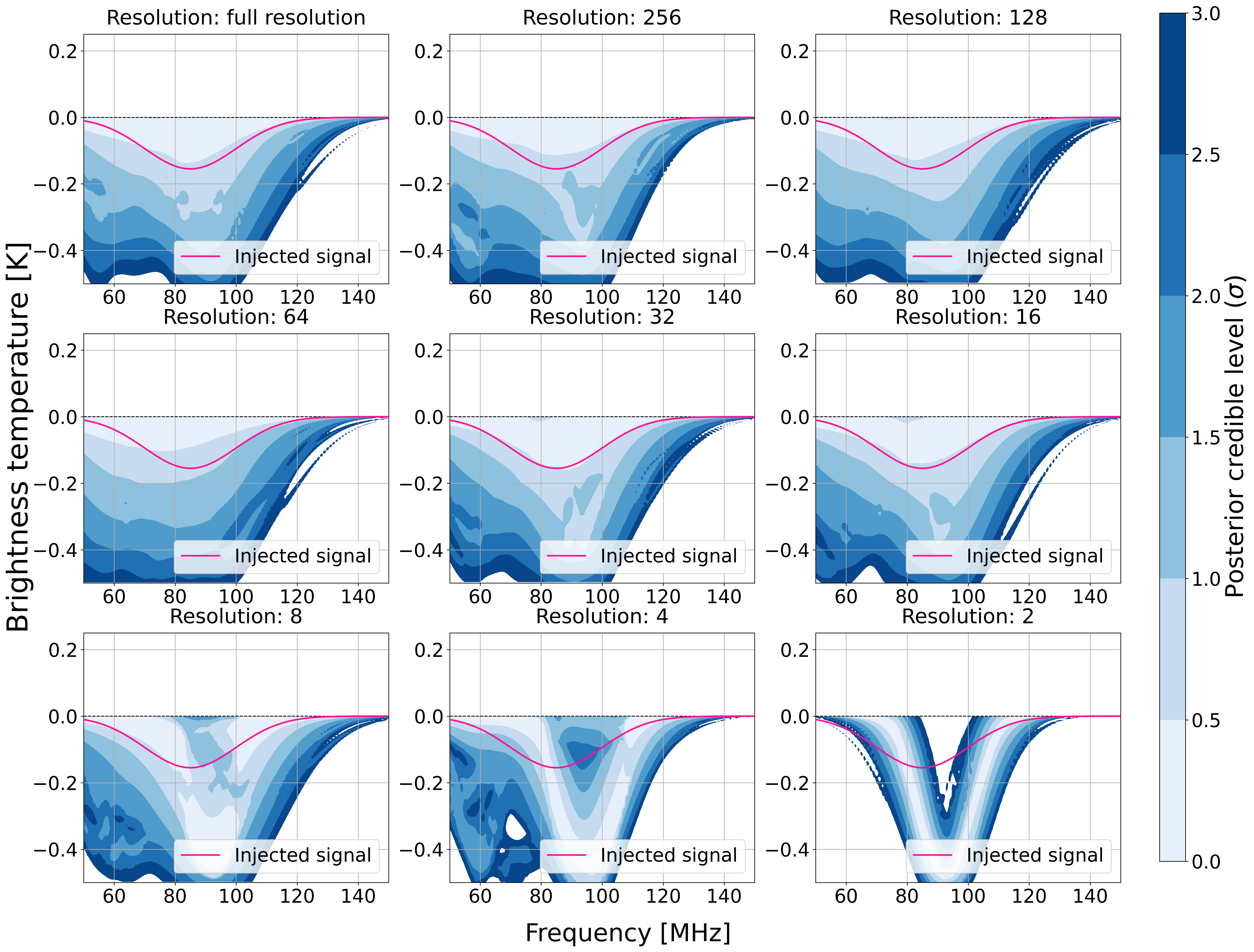}
        \caption{Spatially varying spectral index}
        \label{fig:full_sp_loglog_gauss}
    \end{subfigure}

    \caption{Recovered signal posteriors in comparison to the injected Gaussian. In this case data was fit with the log-log polynomial described in \ref{eqn:loglog_polynomial_intro}}
    \label{fig:loglog_fit_gauss}
\end{figure*}

\begin{figure*}
    \centering

    \begin{subfigure}{0.49\textwidth}
        \centering
        \includegraphics[width=\linewidth]{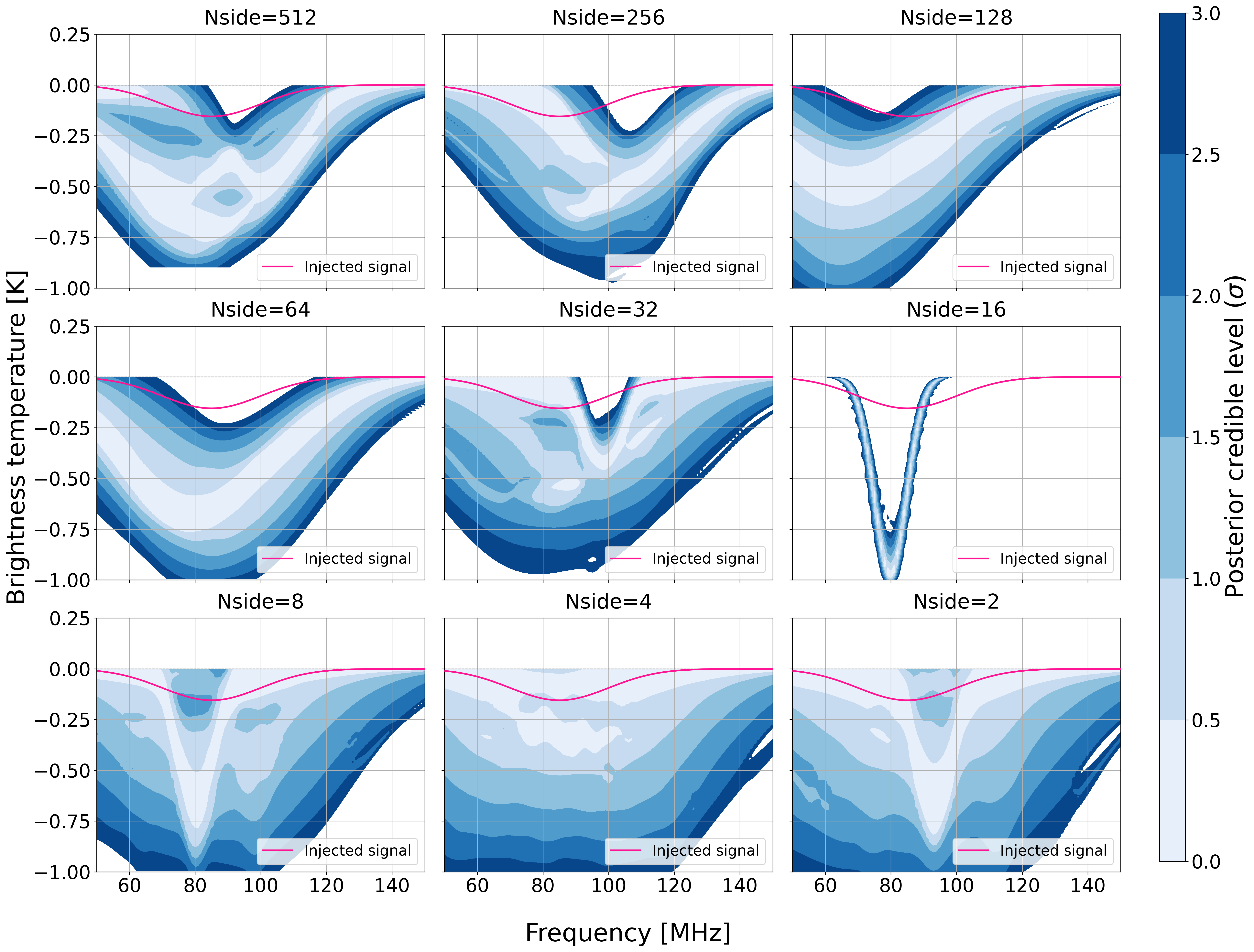}
        \caption{Uniform spectral index}
        \label{fig:uniform_diff_gauss}
    \end{subfigure}
    \hfill
    \begin{subfigure}{0.49\textwidth}
        \centering
        \includegraphics[width=\linewidth]{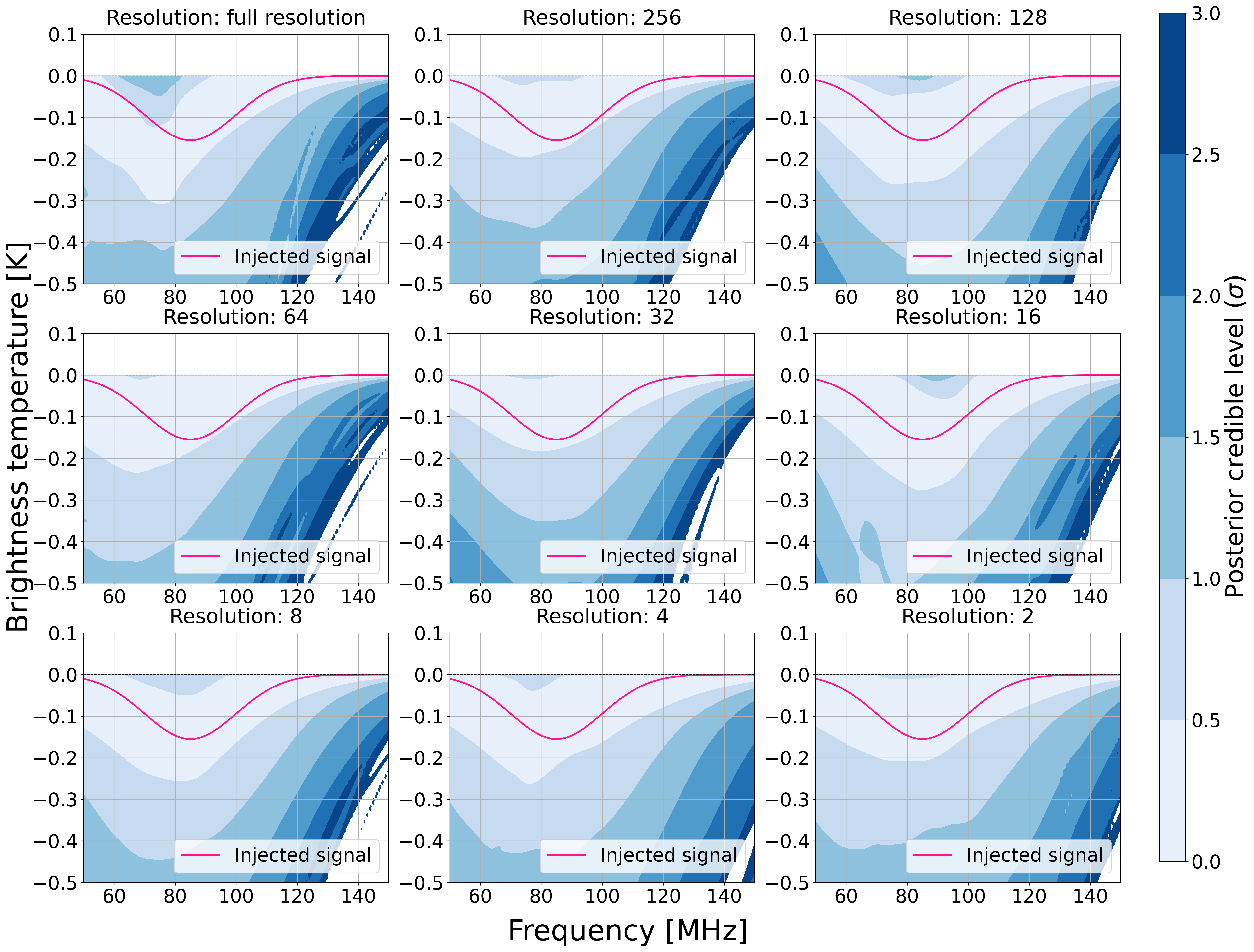}
        \caption{Spatially varying spectral index}
        \label{fig:full_sp_diff_gauss}
    \end{subfigure}

    \caption{Recovered signal posteriors in comparison to the injected Gaussian. In this case data was fit with the difference polynomial described in \ref{eqn:differencepolynomial_intro}}
    \label{fig:diff_fit_gauss}
\end{figure*}

This signal, visually shown in Figure \ref{fig:injectedsignals}, is much shallower and of the same order of magnitude as the chromatic effects introduced by the beam, making it significantly harder to detect. Gaussian white noise is present in all simulated data, drawn from a normal distribution with zero mean and standard deviation $\sigma = 0.025$. The amplitude of the shallow Gaussian is strongly degenerate with the foreground model's own coefficients, which can flex to absorb a signal of this depth regardless of the foreground model used. As a result, retrieving this shallow Gaussian signal did not prove fruitful as seen in Figures \ref{fig:edges_gauss}, \ref{fig:loglog_fit_gauss} and \ref{fig:diff_fit_gauss}.

\section{Conclusion}
\label{sec:conclusion}
In this work, we have investigated the sky map resolution requirements for beam chromaticity correction in sky-averaged 21 cm experiments, using simulated data. Our analysis considered three foreground models: the difference polynomial, the log-log polynomial, and the EDGES-style power-law expansion, applied under both constrained and unconstrained fitting frameworks, with the sky map progressively degraded from slightly blurred to completely blurred without any recognisable features still visible on the sky.

Chromaticity correction residuals remain stable down to map resolutions of $N_{\rm side}=64$ which coincides with the precision error of $2.1\%$ and angular resolution of $\sim 1^\circ$, with all three foreground models and both fitting frameworks showing negligible degradation across this range. The amplitude of residuals begins to increase noticeably only below $N_{\rm side}=32$ ($3.7\%$ divergence from the original map, $\sim 1.8^\circ$), and diverges significantly at $N_{\rm side}=16$ ($\sim 13\%$, angular size of $\sim 3.6^\circ$) and below. This suggests that the sky map does not need to be at full resolution to produce a reliable chromaticity correction, and that moderate degradation minimal change to the foreground residuals.

The choice of foreground model has a larger impact on signal recovery than map resolution. The EDGES power-law expansion, being physically aligned with the spectral structure of Galactic synchrotron emission, requires fewer terms to achieve comparable residuals and yields cleaner signal posteriors than either the difference polynomial or the log-log polynomial. The difference polynomial requires the highest order ($N=15$) under both the unconstrained and constrained framework. The unconstrained fit in Figure \ref{fig:diff_fit_fgauss} produces the widest posterior distributions for signal parameters, suggesting that its additional flexibility allows it to absorb signal power into the foreground fit.

Recovery of the shallow Gaussian signal ($A = 0.155$~K) proved unsuccessful at with all foreground models due to signal parameters being strongly correlated to foreground parameters in our analysis. 

Recovery of the deeper flattened Gaussian signal ($A = 0.5$~K) was partially successful, particularly with the EDGES power-law foreground under a uniform spectral index assumption. Under these conditions, the posterior credible intervals are narrow and centred close to the injected values at high map resolutions. Performance degrades with spatially varying spectral index scaling, which introduces additional chromatic structure that competes with the signal.

These results have practical implications for the design and calibration of global 21 cm experiments. They suggest that residuals in signal recovery are dominated by foreground model selection and noise rather than imperfect sky knowledge. Future work should focus on identifying foreground parameterisations that are both spectrally flexible enough to capture chromatic distortions and constrained enough to avoid absorbing the cosmological signal.

\section*{Acknowledgements}

The authors acknowledge UKRI for their support of this research via the Horizon Europe Guarantee grant (ERC Consolidator) REACH\_21 (EP/Y02916X/1). EdLA acknowledges the support of UKRI STFC via an Ernest Rutherford Fellowship (ST/V004425/1). HTJB acknowledges support from the Kavli Institute for Cosmology Cambridge and the Kavli Foundation. We would also like to thank John Cumner for providing the simulated beam of the REACH dipole antenna and Rohan Patel for helpful feedback on the design and readability of the figures in this work.

\bibliographystyle{mnras}
\bibliography{ref}

@article{Maxsmoothpaper,
    author = {Bevins, H T J and Handley, W J and Fialkov, A and de Lera Acedo, E and Greenhill, L J and Price, D C},
    title = { maxsmooth: rapid maximally smooth function fitting with applications in Global 21-cm cosmology},
    journal = {Monthly Notices of the Royal Astronomical Society},
    volume = {502},
    number = {3},
    pages = {4405-4425},
    year = {2021},
    month = {01},
    issn = {0035-8711},
    doi = {10.1093/mnras/stab152},
    url = {https://doi.org/10.1093/mnras/stab152},
    eprint = {https://academic.oup.com/mnras/article-pdf/502/3/4405/38869834/stab152.pdf},
}

@article{Mozdzen2016,
    author = {Mozdzen, T. J. and Bowman, J. D. and Monsalve, R. A. and Rogers, A. E. E.},
    title = {Improved measurement of the spectral index of the diffuse radio background between 90 and 190 MHz},
    journal = {Monthly Notices of the Royal Astronomical Society},
    volume = {464},
    number = {4},
    pages = {4995-5002},
    year = {2016},
    month = {11},
    issn = {0035-8711},
    doi = {10.1093/mnras/stw2696},
    url = {https://doi.org/10.1093/mnras/stw2696},
    eprint = {https://academic.oup.com/mnras/article-pdf/464/4/4995/8417919/stw2696.pdf},
}

@article{Mozdzen2017,
    author = {Mozdzen, T J and Mahesh, N and Monsalve, R A and Rogers, A E E and Bowman, J D},
    title = {Spectral index of the diffuse radio background between 50 and 100 MHz},
    journal = {Monthly Notices of the Royal Astronomical Society},
    volume = {483},
    number = {4},
    pages = {4411-4423},
    year = {2019},
    month = {03},
    issn = {0035-8711},
    doi = {10.1093/mnras/sty3410},
    url = {https://doi.org/10.1093/mnras/sty3410},
    eprint = {https://academic.oup.com/mnras/article-pdf/483/4/4411/27566434/sty3410.pdf},
}

@article{Dominic2021,
    author = {Anstey, Dominic and de Lera Acedo, Eloy and Handley, Will},
    title = {A general Bayesian framework for foreground modelling and chromaticity correction for global 21 cm experiments},
    journal = {Monthly Notices of the Royal Astronomical Society},
    volume = {506},
    number = {2},
    pages = {2041-2058},
    year = {2021},
    month = {06},
    issn = {0035-8711},
    doi = {10.1093/mnras/stab1765},
    url = {https://doi.org/10.1093/mnras/stab1765},
    eprint = {https://academic.oup.com/mnras/article-pdf/506/2/2041/42467147/stab1765.pdf},
}

@article{Dominic2023,
    author = {Anstey, Dominic and de Lera Acedo, Eloy and Handley, Will},
    title = {Use of time dependent data in Bayesian global 21-cm foreground and signal modelling},
    journal = {Monthly Notices of the Royal Astronomical Society},
    volume = {520},
    number = {1},
    pages = {850-865},
    year = {2023},
    month = {03},
    issn = {0035-8711},
    doi = {10.1093/mnras/stad156},
    url = {https://doi.org/10.1093/mnras/stad156},
    eprint = {https://academic.oup.com/mnras/article-pdf/520/1/850/49057286/stad156.pdf},
}

@article{REACH,
author = {Acedo, Eloy and De Villiers, Dirk and Razavi-Ghods, Nima and Handley, Will and Fialkov, Anastasia and Magro, Alessio and Anstey, D. and Bevins, H. and Chiello, Riccardo and Cumner, J. and Josaitis, A. and Roque, I. and Sims, P. and Scheutwinkel, Kilian and Alexander, Paul and Bernardi, G. and Carey, S. and Cavillot, Jean and Croukamp, W. and Zarb-Adami, K.},
journal = {Nature Astronomy},
number = {6},
year = {2022},
month = {10},
pages = {},
title = {The REACH radiometer for detecting the 21-cm hydrogen signal from redshift 7.5 to 28},
doi = {https://doi.org/10.1038/s41550-022-01709-9}
}

@Article{EDGES2017,
author={Bowman, Judd D.
and Rogers, Alan E. E.
and Monsalve, Raul A.
and Mozdzen, Thomas J.
and Mahesh, Nivedita},
title={An absorption profile centred at 78 megahertz in the sky-averaged spectrum},
journal={Nature},
year={2018},
month={Mar},
day={01},
volume={555},
number={7694},
pages={67-70},
issn={1476-4687},
doi={10.1038/nature25792},
url={https://doi.org/10.1038/nature25792}
}

@ARTICLE{deOliveira-Costa_GSM2008,
       author = {{de Oliveira-Costa}, Ang{\'e}lica and {Tegmark}, Max and {Gaensler}, B.~M. and {Jonas}, Justin and {Landecker}, T.~L. and {Reich}, Patricia},
        title = "{A model of diffuse Galactic radio emission from 10 MHz to 100 GHz}",
      journal = {\mnras},
         year = 2008,
        month = jul,
       volume = {388},
       number = {1},
        pages = {247-260},
          doi = {10.1111/j.1365-2966.2008.13376.x},
archivePrefix = {arXiv},
       eprint = {0802.1525},
 primaryClass = {astro-ph},
       adsurl = {https://ui.adsabs.harvard.edu/abs/2008MNRAS.388..247D}
}

@ARTICLE{Haslam,
       author = {{Haslam}, C.~G.~T. and {Klein}, U. and {Salter}, C.~J. and {Stoffel}, H. and {Wilson}, W.~E. and {Cleary}, M.~N. and {Cooke}, D.~J. and {Thomasson}, P.},
        title = "{A 408 MHz all-sky continuum survey. I - Observations at southern declinations and for the North Polar region.}",
      journal = {\aap},
         year = 1981,
        month = jul,
       volume = {100},
        pages = {209-219},
       adsurl = {https://ui.adsabs.harvard.edu/abs/1981A&A...100..209H}
}

@article{SARAS_disputing_EDGES,
  title     = "On the detection of a cosmic dawn signal in the radio background",
  author    = "Singh, Saurabh and Nambissan, T. Jishnu",
  journal   = "Nat. Astron.",
  publisher = "Springer Science and Business Media LLC",
  volume    =  6,
  number    =  5,
  pages     = "607--617",
  month     =  feb,
  year      =  2022,
  language  = "en"
}

@ARTICLE{Wouthuysen,
       author = {{Wouthuysen}, S.~A.},
        title = "{On the excitation mechanism of the 21-cm (radio-frequency) interstellar hydrogen emission line.}",
      journal = {\aj},
         year = 1952,
        month = jan,
       volume = {57},
        pages = {31-32},
          doi = {10.1086/106661},
       adsurl = {https://ui.adsabs.harvard.edu/abs/1952AJ.....57R..31W}
}

@ARTICLE{Field_a,
       author = {{Field}, George B.},
        title = "{The Spin Temperature of Intergalactic Neutral Hydrogen.}",
      journal = {\apj},
         year = 1959,
        month = may,
       volume = {129},
        pages = {536},
          doi = {10.1086/146653},
       adsurl = {https://ui.adsabs.harvard.edu/abs/1959ApJ...129..536F}
}

@ARTICLE{Field_b,
       author = {{Field}, George B.},
        title = "{The Time Relaxation of a Resonance-Line Profile.}",
      journal = {\apj},
         year = 1959,
        month = may,
       volume = {129},
        pages = {551},
          doi = {10.1086/146654},
       adsurl = {https://ui.adsabs.harvard.edu/abs/1959ApJ...129..551F}
}

@article{SARAS3,
   title={SARAS 3 CD/EoR radiometer: design and performance of the receiver},
   volume={51},
   ISSN={1572-9508},
   url={http://dx.doi.org/10.1007/s10686-020-09697-2},
   DOI={10.1007/s10686-020-09697-2},
   number={2},
   journal={Experimental Astronomy},
   publisher={Springer Science and Business Media LLC},
   author={Nambissan T., Subrahmanyan, Ravi and Somashekar, R. and Shankar, N. Udaya and Singh, Saurabh and Raghunathan, A. and Girish, B. S. and Srivani, K. S. and Rao, Mayuri Sathyanarayana},
   year={2021},
   month=jan, pages={193–234} }

@article{Bayesian_calibration_2023,
   title={A Bayesian method to mitigate the effects of unmodelled time-varying systematics for 21-cm cosmology experiments},
   volume={527},
   ISSN={1365-2966},
   url={http://dx.doi.org/10.1093/mnras/stad3725},
   DOI={10.1093/mnras/stad3725},
   number={3},
   journal={Monthly Notices of the Royal Astronomical Society},
   publisher={Oxford University Press (OUP)},
   author={Kirkham, Christian J and Anstey, Dominic J and de Lera Acedo, Eloy},
   year={2023},
   month=nov, pages={8305–8315} }

@ARTICLE{MSF_2017,
       author = {{Sathyanarayana Rao}, Mayuri and {Subrahmanyan}, Ravi and {Udaya Shankar}, N. and {Chluba}, Jens},
        title = "{Modeling the Radio Foreground for Detection of CMB Spectral Distortions from the Cosmic Dawn and the Epoch of Reionization}",
      journal = {\apj},
         year = 2017,
        month = may,
       volume = {840},
       number = {1},
          eid = {33},
        pages = {33},
          doi = {10.3847/1538-4357/aa69bd},
archivePrefix = {arXiv},
       eprint = {1611.04602},
 primaryClass = {astro-ph.CO},
       adsurl = {https://ui.adsabs.harvard.edu/abs/2017ApJ...840...33S}
}

@ARTICLE{MSF_2015,
       author = {{Sathyanarayana Rao}, Mayuri and {Subrahmanyan}, Ravi and {Udaya Shankar}, N. and {Chluba}, Jens},
        title = "{On the Detection of Spectral Ripples from the Recombination Epoch}",
      journal = {\apj},
         year = 2015,
        month = sep,
       volume = {810},
       number = {1},
          eid = {3},
        pages = {3},
          doi = {10.1088/0004-637X/810/1/3},
archivePrefix = {arXiv},
       eprint = {1501.07191},
 primaryClass = {astro-ph.IM},
       adsurl = {https://ui.adsabs.harvard.edu/abs/2015ApJ...810....3S}
}

@article{nested_sampling,
author = {John Skilling},
title = {{Nested sampling for general Bayesian computation}},
volume = {1},
journal = {Bayesian Analysis},
number = {4},
publisher = {International Society for Bayesian Analysis},
pages = {833 -- 859},
year = {2006},
doi = {10.1214/06-BA127},
URL = {https://doi.org/10.1214/06-BA127}
}

@ARTICLE{REACH_antenna_design,
       author = {{Cumner}, J. and {de Lera Acedo}, E. and {de Villiers}, D.~I.~L. and {Anstey}, D. and {Kolitsidas}, C.~I. and {Gurdon}, B. and {Fagnoni}, N. and {Alexander}, P. and {Bernardi}, G. and {Bevins}, H.~T.~J. and {Carey}, S. and {Cavillot}, J. and {Chiello}, R. and {Craeye}, C. and {Croukamp}, W. and {Ely}, J.~A. and {Fialkov}, A. and {Gessey-Jones}, T. and {Gueuning}, Q. and {Handley}, W. and {Hills}, R. and {Josaitis}, A.~T. and {Kulkarni}, G. and {Magro}, A. and {Maiolino}, R. and {Meerburg}, P.~D. and {Mittal}, S. and {Pritchard}, J.~R. and {Puchwein}, E. and {Razavi-Ghods}, N. and {Roque}, I.~L.~V. and {Saxena}, A. and {Scheutwinkel}, K.~H. and {Shen}, E. and {Sims}, P.~H. and {Smirnov}, O. and {Spinelli}, M. and {Zarb-Adami}, K.},
        title = "{Radio Antenna Design for Sky-Averaged 21cm Cosmology Experiments: The REACH Case}",
      journal = {Journal of Astronomical Instrumentation},
         year = 2022,
        month = jan,
       volume = {11},
       number = {1},
          eid = {2250001-2058},
        pages = {2250001-2058},
          doi = {10.1142/S2251171722500015},
archivePrefix = {arXiv},
       eprint = {2109.10098},
 primaryClass = {astro-ph.IM},
       adsurl = {https://ui.adsabs.harvard.edu/abs/2022JAI....1150001C}
}

@ARTICLE{REACH_bayesian_pipeline_dominic,
       author = {{Anstey}, Dominic and {Cumner}, John and {de Lera Acedo}, Eloy and {Handley}, Will},
        title = "{Informing antenna design for sky-averaged 21-cm experiments using a simulated Bayesian data analysis pipeline}",
      journal = {\mnras},
         year = 2022,
        month = feb,
       volume = {509},
       number = {4},
        pages = {4679-4693},
          doi = {10.1093/mnras/stab3211},
archivePrefix = {arXiv},
       eprint = {2106.10193},
 primaryClass = {astro-ph.IM},
       adsurl = {https://ui.adsabs.harvard.edu/abs/2022MNRAS.509.4679A}
}

@article{Cohen_Fialkov_2017,
    author = {Cohen, Aviad and Fialkov, Anastasia and Barkana, Rennan and Lotem, Matan},
    title = {Charting the parameter space of the global 21-cm signal},
    journal = {Monthly Notices of the Royal Astronomical Society},
    volume = {472},
    number = {2},
    pages = {1915-1931},
    year = {2017},
    month = {08},
    issn = {0035-8711},
    doi = {10.1093/mnras/stx2065},
    url = {https://doi.org/10.1093/mnras/stx2065},
    eprint = {https://academic.oup.com/mnras/article-pdf/472/2/1915/19917726/stx2065.pdf},
}

@article{Cohen_Fialkov_2020,
    author = {Cohen, Aviad and Fialkov, Anastasia and Barkana, Rennan and Monsalve, Raul A},
    title = {Emulating the global 21-cm signal from Cosmic Dawn and Reionization},
    journal = {Monthly Notices of the Royal Astronomical Society},
    volume = {495},
    number = {4},
    pages = {4845-4859},
    year = {2020},
    month = {06},
    issn = {0035-8711},
    doi = {10.1093/mnras/staa1530},
    url = {https://doi.org/10.1093/mnras/staa1530},
    eprint = {https://academic.oup.com/mnras/article-pdf/495/4/4845/33382403/staa1530.pdf},
}

@article{FengHolder2018,
  author = {Feng, Chang and Holder, Gilbert P.},
  title = {Enhanced global signal of neutral hydrogen due to excess radiation at cosmic dawn},
  journal = {The Astrophysical Journal Letters},
  year = {2018},
  volume = {858},
  number = {2},
  pages = {L17},
  doi = {10.3847/2041-8213/aac0fe}
}

@article{Modeling_from_edges,
  author = {Ewall-Wice, Aaron and Chang, Tzu-Ching and Lazio, Joseph and Dore, Olivier and Seiffert, Michael and Monsalve, Raul A. and Bowman, Judd D. and Rogers, Alan E. E.},
  title = {Modeling the Radio Background from Early Galaxies in Relation to the EDGES 21 cm Absorption Signal},
  journal = {The Astrophysical Journal},
  year = {2018},
  volume = {868},
  number = {1},
  pages = {63},
  doi = {10.3847/1538-4357/aae51d}
}

@article{Pober_Sims_2019,
    author = {Sims, Peter H and Pober, Jonathan C},
    title = {Testing for calibration systematics in the EDGES low-band data using Bayesian model selection},
    journal = {Monthly Notices of the Royal Astronomical Society},
    volume = {492},
    number = {1},
    pages = {22-38},
    year = {2019},
    month = {12},
    issn = {0035-8711},
    doi = {10.1093/mnras/stz3388},
    url = {https://doi.org/10.1093/mnras/stz3388},
    eprint = {https://academic.oup.com/mnras/article-pdf/492/1/22/31906780/stz3388.pdf},
}

@article{excess_radio_background_fialkov,
    author = {Fialkov, Anastasia and Barkana, Rennan},
    title = {Signature of excess radio background in the 21-cm global signal and power spectrum},
    journal = {Monthly Notices of the Royal Astronomical Society},
    volume = {486},
    number = {2},
    pages = {1763-1773},
    year = {2019},
    month = {03},
    issn = {0035-8711},
    doi = {10.1093/mnras/stz873},
    url = {https://doi.org/10.1093/mnras/stz873},
    eprint = {https://academic.oup.com/mnras/article-pdf/486/2/1763/28484631/stz873.pdf},
}

@article{primordial_black_holes,
  title = {Cosmic radio background from primordial black holes at cosmic dawn},
  author = {Zhang, Zhihe and Yue, Bin and Xu, Yidong and Ma, Yin-Zhe and Chen, Xuelei and Liu, Maoyuan},
  journal = {Phys. Rev. D},
  volume = {107},
  issue = {8},
  pages = {083013},
  numpages = {16},
  year = {2023},
  month = {Apr},
  publisher = {American Physical Society},
  doi = {10.1103/PhysRevD.107.083013},
  url = {https://link.aps.org/doi/10.1103/PhysRevD.107.083013}
}

@article{Hills_2018,
   title={Concerns about modelling of the EDGES data},
   volume={564},
   ISSN={1476-4687},
   url={http://dx.doi.org/10.1038/s41586-018-0796-5},
   DOI={10.1038/s41586-018-0796-5},
   number={7736},
   journal={Nature},
   publisher={Springer Science and Business Media LLC},
   author={Hills, Richard and Kulkarni, Girish and Meerburg, P. Daniel and Puchwein, Ewald},
   year={2018},
   month=Dec, pages={E32–E34} }

@article{Singh_2019_sunisoidal_edges_feature,
   title={The Redshifted 21 cm Signal in the EDGES Low-band Spectrum},
   volume={880},
   ISSN={1538-4357},
   url={http://dx.doi.org/10.3847/1538-4357/ab2879},
   DOI={10.3847/1538-4357/ab2879},
   number={1},
   journal={The Astrophysical Journal},
   publisher={American Astronomical Society},
   author={Singh, Saurabh and Subrahmanyan, Ravi},
   year={2019},
   month=July, pages={26} }

@article{Pagano_bayesian_modeling,
    author = {Pagano, Michael and Sims, Peter and Liu, Adrian and Anstey, Dominic and Handley, Will and de Lera Acedo, Eloy},
    title = {A general Bayesian framework to account for foreground map errors in global 21-cm experiments},
    journal = {Monthly Notices of the Royal Astronomical Society},
    volume = {527},
    number = {3},
    pages = {5649-5667},
    year = {2024},
    month = {01},
    issn = {0035-8711},
    doi = {10.1093/mnras/stad3392},
    url = {https://doi.org/10.1093/mnras/stad3392},
    eprint = {https://academic.oup.com/mnras/article-pdf/527/3/5649/53980125/stad3392.pdf},
}

@article{emma_ionospheric_effects,
    author = {Shen, Emma and Anstey, Dominic and de Lera Acedo, Eloy and Fialkov, Anastasia and Handley, Will},
    title = {Quantifying ionospheric effects on global 21-cm observations},
    journal = {Monthly Notices of the Royal Astronomical Society},
    volume = {503},
    number = {1},
    pages = {344-353},
    year = {2021},
    month = {05},
    issn = {0035-8711},
    doi = {10.1093/mnras/stab429},
    url = {https://doi.org/10.1093/mnras/stab429},
    eprint = {https://academic.oup.com/mnras/article-pdf/503/1/344/38845082/stab429.pdf},
}

@article{fgivenx,
    doi = {10.21105/joss.00849},
    url = {https://doi.org/10.21105/joss.00849},
    year = {2018},
    month = {Aug},
    publisher = {The Open Journal},
    volume = {3},
    number = {28},
    pages = {849},
    author = {Handley, Will},
    title = {fgivenx: A Python package for functional posterior plotting},
    journal = {Journal of Open Source Software}
}

@misc{Cabezas2024,
        title={BlackJAX: Composable Bayesian inference in JAX},
        author={Alberto Cabezas and Adrien Corenflos and Junpeng Lao and Rémi Louf},
        year={2024},
        eprint={2402.10797},
        archivePrefix={arXiv},
        primaryClass={cs.MS}
}

@software{jax2018github,
  author = {James Bradbury and Roy Frostig and Peter Hawkins and Matthew James Johnson and Chris Leary and Dougal Maclaurin and George Necula and Adam Paszke and Jake Vander{P}las and Skye Wanderman-{M}ilne and Qiao Zhang},
  title = {{JAX}: composable transformations of {P}ython+{N}um{P}y programs},
  url = {http://github.com/jax-ml/jax},
  version = {0.3.13},
  year = {2018},
}

@unknown{Jacobs_pipeline_paper_2026,
author = {Tutt, Jacob and Sims, Peter and Pattison, Joe and Anstey, Dominic and Leeney, Samuel and Acedo, Eloy},
year = {2026},
month = {03},
pages = {},
title = {Optimising Foreground Modelling for Global 21cm Cosmology with GPU-Accelerated Nested Sampling},
doi = {10.48550/arXiv.2603.13196}
}

@unknown{signatures_of_cr_heating_tomas,
author = {Gessey-Jones, T. and Fialkov, Anastasia and Acedo, Eloy and Handley, Will and Barkana, R.},
year = {2023},
month = {04},
pages = {},
title = {Signatures of Cosmic Ray Heating in 21-cm Observables},
doi = {10.48550/arXiv.2304.07201}
}

@article{nested_sampling_yallup_gpu,
title={Nested Slice Sampling: Vectorized Nested Sampling for {GPU}-Accelerated Inference},
author={David Yallup and Namu Kroupa and Will Handley},
journal={Transactions on Machine Learning Research},
issn={2835-8856},
year={2026},
url={https://openreview.net/forum?id=5mF2eRl3gt},
note={}
}

@article{msf_proposition_2017,
doi = {10.3847/1538-4357/aa69bd},
url = {https://doi.org/10.3847/1538-4357/aa69bd},
year = {2017},
month = {may},
publisher = {The American Astronomical Society},
volume = {840},
number = {1},
pages = {33},
author = {Sathyanarayana Rao, Mayuri and Subrahmanyan, Ravi and Shankar, N Udaya and Chluba, Jens},
title = {Modeling the Radio Foreground for Detection of CMB Spectral Distortions from the Cosmic Dawn and the Epoch of Reionization},
journal = {The Astrophysical Journal},
}

@article{msf_proposition_2015,
doi = {10.1088/0004-637X/810/1/3},
url = {https://doi.org/10.1088/0004-637X/810/1/3},
year = {2015},
month = {aug},
publisher = {The American Astronomical Society},
volume = {810},
number = {1},
pages = {3},
author = {Sathyanarayana Rao, Mayuri and Subrahmanyan, Ravi and Shankar, N Udaya and Chluba, Jens},
title = {ON THE DETECTION OF SPECTRAL RIPPLES FROM THE RECOMBINATION EPOCH},
journal = {The Astrophysical Journal}
}

@ARTICLE{healpix,
       author = {{G{\'o}rski}, K.~M. and {Hivon}, E. and {Banday}, A.~J. and {Wandelt}, B.~D. and {Hansen}, F.~K. and {Reinecke}, M. and {Bartelmann}, M.},
        title = "{HEALPix: A Framework for High-Resolution Discretization and Fast Analysis of Data Distributed on the Sphere}",
      journal = {\apj},
         year = 2005,
        month = apr,
       volume = {622},
       number = {2},
        pages = {759-771},
          doi = {10.1086/427976},
archivePrefix = {arXiv},
       eprint = {astro-ph/0409513},
 primaryClass = {astro-ph},
       adsurl = {https://ui.adsabs.harvard.edu/abs/2005ApJ...622..759G}
}

\bsp
\label{lastpage}
\end{document}